# Steklov Spectral Geometry for Extrinsic Shape Analysis

YU WANG, Massachusetts Institute of Technology, USA
MIRELA BEN-CHEN, Technion – Israel Institute of Technology, Israel
IOSIF POLTEROVICH, Université de Montréal, Canada
JUSTIN SOLOMON, Massachusetts Institute of Technology, USA

We propose using the *Dirichlet-to-Neumann operator* as an extrinsic alternative to the Laplacian for spectral geometry processing and shape analysis. Intrinsic approaches, usually based on the Laplace–Beltrami operator, cannot capture the spatial embedding of a shape up to rigid motion, and many previous extrinsic methods lack theoretical justification. Instead, we consider the Steklov eigenvalue problem, computing the spectrum of the Dirichlet-to-Neumann operator of a surface bounding a volume. A remarkable property of this operator is that it completely encodes volumetric geometry. We use the boundary element method (BEM) to discretize the operator, accelerated by hierarchical numerical schemes and preconditioning; this pipeline allows us to solve eigenvalue and linear problems on large-scale meshes despite the density of the Dirichlet-to-Neumann discretization. We further demonstrate that our operators naturally fit into existing frameworks for geometry processing, making a shift from intrinsic to extrinsic geometry as simple as substituting the Laplace–Beltrami operator with the Dirichlet-to-Neumann operator.

## 1 INTRODUCTION

Geometry processing and shape analysis tools for computer graphics typically draw from two complementary theories of geometry. To distinguish them, consider a closed surface embedded in three dimensions. From the viewpoint of *extrinsic geometry*, we might examine the surface as the outer boundary of a volume. This approach relies on distances and other measurements taken from the space surrounding the surface to understand its shape. Contrastingly, many techniques in differential geometry decouple extrinsic shape from *intrinsic geometry*, which is concerned with quantities like geodesic distances that can be measured without leaving the outer surface. A crowning achievement of classical differential geometry shows that certain quantities like Gaussian curvature can be measured intrinsically; this basic observation inspired theoretical exploration of purely intrinsic techniques. Computational geometry processing embraced the intrinsic perspective early on, leading to numerous applications of intrinsic computations.

Intrinsic geometry, however, is an ineffective description of shape for many applications. First, the spatial embedding information is lost. The variability in embedding can be essential. As an extreme example, consider searching a database of origami models: From an intrinsic perspective, all origami is equivalent to the same piece of flat paper. Second, the intrinsic perspective is a counterintuitive way to describe the shapes of many real-world objects, e.g. identifying the inward and outward bumps on the cubes in Figure 1.

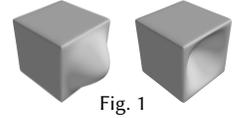

Fig. 1

A naïve approach to extrinsic geometry could be to use the $(x, y, z)$ coordinates of the embedding. This incorporates information about the embedding but is not invariant to rigid motion. Alternatively, rotation-invariant shape descriptors (e.g., built from spherical harmonic power spectra) usually involve extrinsic information about a surface as a shell rather than as the boundary of a volume.

In this paper, we provide a practical and mathematically-justified spectral approach to extrinsic geometry for geometry processing, via an extrinsic alternative to the intrinsic Laplace–Beltrami operator. The end result is a *surface-only* approach to *volume-aware* shape analysis, using a boundary operator that takes the interaction between non-adjacent vertices into consideration. In particular, we consider the Dirichlet-to-Neumann (DtN) operator, also known as the Poincaré-Steklov or simply the Steklov operator, and its spectrum (the *Steklov spectrum*). A physical interpretation of the DtN operator is that it maps from a temperature distribution on the surface to the resulting heat flux through the surface. As we will prove, DtN and its spectrum completely encode extrinsic geometry in the smooth case. We also show that DtN and some of its peers can be applied efficiently using the boundary element method (BEM), accelerated using preconditioning and hierarchical techniques that scale linearly in the size of a boundary triangle mesh.

We generalize our method to additional geometric primitives, notably including surfaces with open boundary. Since it is built using the Laplace equation, our discretized Dirichlet-to-Neumann operator naturally fits into existing frameworks for geometry processing and analysis. We show that in applications including compression, measurement of shape differences, and spectral distance computation, shifting from intrinsic geometry to extrinsic geometry is as simple as substituting the cotangent Laplacian with a discretized

Author's addresses: Y. Wang and J. Solomon, Computer Science and Artificial Intelligence Laboratory, Electrical Engineering and Computer Science, Massachusetts Institute of Technology (address: 32 Vassar St, Cambridge, MA 02139, USA); I. Polterovich, Département de mathématiques et de statistique, Université de Montréal; Mirela Ben-Chen, Computer Science Department, Technion – Israel Institute of Technology.

Authors' addresses: Yu Wang, Massachusetts Institute of Technology, 32 Vassar St, Cambridge, MA, 02139, USA; Mirela Ben-Chen, Technion – Israel Institute of Technology, Israel; Iosif Polterovich, Université de Montréal, Département de mathématiques et de statistique, Montréal, QC, H3C 3J7, Canada; Justin Solomon, Massachusetts Institute of Technology, 32 Vassar St, Cambridge, MA, 02139, USA.

Steklov operator. This simple change leads to qualitatively different behavior for geometric algorithms.

## 2 RELATED WORK

### 2.1 Intrinsic geometry

Intrinsic computations are ubiquitous in geometry processing, applied to problems including distance computation [Crane et al. 2013; Lipman et al. 2010], segmentation [Reuter et al. 2009], shape description [Kokkinos et al. 2012], shape retrieval [Bronstein et al. 2011], and correspondence [Ovsjanikov et al. 2012]; these citations are a small sampling of the literature.

Many intrinsic algorithms use the Laplace–Beltrami operator, which can be approximated e.g. with the celebrated cotangent formula [Pinkall and Polthier 1993] or with convergent approximations on simplicial complexes [Belkin et al. 2008] and point clouds [Belkin et al. 2009; Liang et al. 2012]. Laplace–Beltrami eigenfunctions are multiscale and can characterize shape up to isometry [Aubry et al. 2011; Sun et al. 2009]; truncation in this basis also provides favorable approximation properties [Aflalo et al. 2015]. See the surveys by Zhang et al. [2010], Sorkine [2005], and Patané [2016] for discussion of this technique and applications.

Functional maps, proposed by Ovsjanikov et al. [2012], have inspired interest in a broader "operator-based" approach to geometry. Operator-based geometry processing now includes intrinsic techniques for shape exploration [Rustamov et al. 2013], vector field processing [Azencot et al. 2013, 2015], interpolation [Azencot et al. 2016], simulation [Azencot et al. 2014], and deformation [Boscaini et al. 2015]—among other tasks.

The intrinsic approach is desirable in many scenarios. Many real world objects deform near-isometrically, that is, without affecting intrinsic structure. Isometry invariance then creates an advantage: An object can be bent to another pose, while still being considered the same intrinsically. For example, intrinsic invariance can be favorable in shape retrieval since the same object in different poses will be clustered together by design.

### 2.2 Extrinsic geometry

Extrinsic techniques appear in geometry processing, usually without the multiscale and stability advantages of spectral geometry. The most common expression of discrete extrinsic geometry is in computation of mean curvature, the extrinsic counterpart of Gaussian curvature. Extrinsic methods appear in computer vision, e.g. the SHOT [Tombari et al. 2010] and D2 [Osada et al. 2002] descriptors, typically without completeness guarantees.

A few methods attempt to extend operator-based methods to incorporate extrinsic information. The volumetric heat kernel signature [Raviv et al. 2010] uses a meshing of the volume enclosed by a surface, applied to segmentation by Litman et al. [2012]; Wang and Wang [2015] present a similar construction on tetrahedral meshes. This is computationally expensive and has the property that results depend on the resolution of the interior mesh; note that methods like [Patané 2015] may accelerate heat kernel evaluation but still require meshing of the interior. Hildebrandt et al. [2010] consider a family of deformation energies whose Hessian eigenmodes are sensitive to extrinsic features. Rustamov [2011] proposes extending surface-based Laplace–Beltrami eigenfunctions to the interior of a domain using generalized barycentric coordinates. This extends descriptors like the heat kernel signature (HKS) from the outer surface to the interior but is still intrinsic in the sense that when it is restricted to the outer surface it coincides with the surface-based computation. Wang et al. [2014] modify the Laplace–Beltrami operator to include concavity information, aiding in applications like segmentation without distinguishing between the interior and exterior of a closed surface. Similarly, Corman et al. [2017] use the Laplacian of a meshed shell around a triangulated surface, incorporating mean curvature but treating the surface as a thin shell rather than the boundary of a volume. Techniques like [Chuang et al. 2009] use extrinsic calculations to approximate the Laplace–Beltrami operator; these methods aim to capture intrinsic geometry.

We use the Dirichlet-to-Neumann (DtN) operator as an extrinsic analog to the Laplace–Beltrami operator. In graphics, Gao et al. [2014] use DtN for skinning; it also appears in a recent pipeline for parameterization [Sawhney and Crane 2017]. In simulation and PDEs, it appears as a Schur complement eliminating interior degrees of freedom [Bertoluzza 2003; Liu et al. 2016; Quarteroni and Valli 1999; Smith et al. 2004; Toselli and Widlund 2005].

Recent concurrent work proposes using the Dirac operator to capture extrinsic shape [Liu et al. 2017]. This method is among the first to extend spectral geometry processing to include extrinsic information, but it still treats the surface as a shell rather than as the boundary of a volume. While this work opens intriguing theoretical questions regarding the informativeness of the Dirac operator (see their §7), the Dirichlet-to-Neumann operator we propose for shape analysis enjoys strong grounding in existing theory; this allows us to prove in §3.1 that the Dirichlet-to-Neumann operator fully captures extrinsic geometry up to rigid motion (Proposition 3.1). On the other hand, unlike the operators we consider in this paper, their Dirac operator benefits from the efficiency of sparse linear algebra. See §7.7 for additional discussion and empirical comparison to this technique.

### 2.3 Numerical PDE

Two typical numerical methods for approximating Dirichlet-to-Neumann operators are the finite element method (FEM) and the boundary element method (BEM). FEM yields a sparse linear system on a discretization of the volume bounded by a surface, while BEM yields a dense linear system on just the boundary. Efficiency-wise, the added number of elements required for FEM makes the two methods similar asymptotically. We use BEM to avoid dependence on volumetric meshing.

Fast numerical methods can be used to accelerate BEM, including the fast multipole method (FMM) and the hierarchical matrix (H-matrix) method. FMM [Engheta et al. 1992; Rokhlin 1985] expands the Green's function using a multipole approximation, grouping close points as a single source. The H-matrix method represents BEM operators hierarchically, reducing the cost of assembly and matrix-vector products to $O(n \log n)$ time [Börm et al. 2003; Hackbusch 1999]. Either method reduces evaluation cost to $O(n \log 1/\epsilon)$ time, where $\epsilon$ is a prescribed accuracy, making it possible to solve large problems on a single workstation.

## 3 MATHEMATICAL PRELIMINARIES

### 3.1 Steklov eigenproblem

Denote a volumetric domain as $\Omega \subseteq \mathbb{R}^3$ with boundary surface $\Gamma = \partial\Omega$. The Steklov eigenvalue problem is defined as:

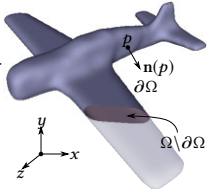

$$\begin{cases} \Delta \psi(x) = 0 & x \in \Omega \\ \nabla_{\mathbf{n}} \psi(x) = \lambda \psi(x) & x \in \Gamma \end{cases}$$

where $\Delta = \partial^2/\partial x^2 + \partial^2/\partial y^2 + \partial^2/\partial z^2$ denotes the Laplacian operator and $\nabla_{\mathbf{n}}$ the normal derivative at the boundary. The spectrum of the Steklov problem is discrete and given by a sequence of eigenvalues

$$0 = \lambda_0 \leq \lambda_1 \leq \lambda_2 \leq \cdots \to \infty.$$

$\{\lambda_n\}_{n=0}^{\infty}$ can be interpreted as the eigenvalues of the Dirichlet-to-Neumann (DtN) operator $\mathcal{S} : H^{1/2}(\Gamma) \to H^{-1/2}(\Gamma)$ defined as $\mathcal{S}f := \nabla_{\mathbf{n}}(\mathcal{E}f)$, where $H^s$ denotes the Sobolev space of order $s$ (see e.g. [Steinbach 2007, §2.4] or [Brezis 2010] for more details). $\mathcal{E}$ denotes the harmonic extension into the interior; denote $\phi_n = \psi_n|_\Gamma : \Gamma \to \mathbb{R}$ as the corresponding eigenfunctions. $\mathcal{S}$ is known as the Dirichlet-to-Neumann operator because it maps the boundary Dirichlet data to the Neumann data (e.g. voltage-to-current, or temperature-to-flux).

Just as the Laplace–Beltrami spectrum encodes intrinsic geometry, the spectrum of $\mathcal{S}$ encodes extrinsic information. In particular, for smooth domains in $\mathbb{R}^3$, the Steklov heat kernel admits the asymptotic expansion as $t \to 0^+$:

$$e^{-t\mathcal{S}}(x, x) = \sum_{i=0}^{\infty} e^{-t\lambda_i} \psi_i(x)^2 \sim \sum_{k=0}^{\infty} a_k(x) t^{k-2} + \sum_{l=1}^{\infty} b_l(x) t^l \log t, \tag{1}$$

where the coefficients $a_k(x)$ for $k = 0, 1, 2$ and $b_l(x)$ are local geometric invariants, in the sense that they are determined by the local geometry of $\Gamma$ in a neighborhood of the point $x \in \Gamma$ [Polterovich and Sher 2015]. See [Duistermaat and Guillemin 1975] where such an expansion is proved for a general elliptic pseudodifferential operator.

The following expressions were obtained by Polterovich and Sher [2015]:

$$a_0(x) \equiv \frac{1}{2\pi} \tag{2}$$

$$a_1(x) = \frac{H(x)}{4\pi} \tag{3}$$

$$a_2(x) = \frac{1}{16\pi} \left( H(x)^2 + \frac{K(x)}{3} \right), \tag{4}$$

where $H(x)$ and $K(x)$ are, respectively, the mean and the Gaussian curvatures of $\Gamma$. While the heat expansion for the Laplace–Beltrami operator on the boundary captures only intrinsic geometric invariants of $\Gamma$, the first few terms of the Steklov heat expansion (1) contain the mean curvature of the boundary, which is an extrinsic geometric quantity. We refer to the recent survey by Girouard and Polterovich [2017] for other results on spectral geometry of $\mathcal{S}$.

Recall that the Laplace–Beltrami operator is invariant under isometries. For example, the bumpy cubes in Figure 1 have isometric boundaries, and therefore their boundary Laplacians coincide. At the same time, the boundaries of the bumpy cubes have different extrinsic geometries, and one could check that the DtN operators on these cubes are also different. Moreover, in striking contrast with the boundary Laplacian, under some assumptions the only maps preserving the DtN operator for domains in $\mathbb{R}^3$ are **rigid motions**:

PROPOSITION 3.1. *Suppose* $\Omega_1, \Omega_2 \subseteq \mathbb{R}^3$ *are compact domains with* $C^\infty$ *boundaries* $\Gamma_1, \Gamma_2$ *and Dirichlet-to-Neumann operators* $\mathcal{S}_1$ *and* $\mathcal{S}_2$, *respectively. Let* $\alpha : \Omega_1 \to \Omega_2$ *be a bijection which is* $C^\infty$ *up to the boundary, and let* $\tilde{\alpha} : \Gamma_1 \to \Gamma_2$ *be the induced mapping between the boundaries. Suppose that the operators* $\mathcal{S}_1$ *and* $\mathcal{S}_2$ *coincide up to composition with* $\tilde{\alpha}$, *i.e.* $\mathcal{S}_2 f = \tilde{\alpha}_* \mathcal{S}_1 \tilde{\alpha}^* f$, *for any* $f \in C^\infty(\Gamma_2)$, *where* $\tilde{\alpha}^* f = f \circ \tilde{\alpha}$, $\tilde{\alpha}_* g = g \circ \tilde{\alpha}^{-1}$ *denote the pull-backs by* $\tilde{\alpha}$ *and* $\tilde{\alpha}^{-1}$, *respectively. Then* $\alpha$ *is a rigid motion.*

PROOF. The following argument was communicated to us by M. Karpukhin: By Proposition 1.3 and the discussion above it in [Lee and Uhlmann 1989], the Dirichlet-to-Neumann map of a smooth surface determines the full Taylor series of the metric $g$ at the boundary in the boundary normal coordinates. In particular, it determines the metric itself—which in our case is the first fundamental form of the boundary surface—and its first derivative, which gives us the second fundamental form (see e.g., [Kachalov et al. 2001, §2.1.18]). At the same time, it follows from Bonnet's theorem that the first and second fundamental forms determine a surface in $\mathbb{R}^3$ up to rigid motions. □

Proving an analogous result for non-smooth boundaries remains a significant challenge for future work.

### 3.2 Boundary representation and operators

To derive a boundary representation of the DtN operator, first consider the Laplace equation with Dirichlet boundary conditions:

$$\begin{cases} \Delta u(x) = 0 & x \in \Omega \\ u(x) = g(x) & x \in \Gamma \end{cases} \tag{5}$$

where $g(\Gamma)$ is Dirichlet data on the boundary. A basic fact from elliptic PDE is that (5) uniquely determines $u(\Omega)$ and hence its Neumann data $g_n = \frac{\partial}{\partial n} u(\Gamma)$. By definition, the Dirichlet-to-Neumann operator $\mathcal{S}$ is the map $g \mapsto g_n$.

To avoid discretizing the interior of the domain $\Omega$, we use integral operators to bypass solving the Laplace equation (5). Here, we introduce relevant boundary operators and notation that will be used throughout the paper. We refer readers to [Steinbach 2007] for a comprehensive introduction to these operators in the context of the boundary element method.

*Single layer potential.* The single layer potential $\mathcal{V} : H^{-1/2}(\Gamma) \to H^{1/2}(\Gamma)$ is defined via

$$[\mathcal{V}\phi](\mathbf{x}) := \int_\Gamma G(\mathbf{x}, \mathbf{y})\phi(\mathbf{y}) \, d\Gamma(\mathbf{y}),$$

where $G(\mathbf{x}, \mathbf{y}) = \frac{1}{4\pi} \frac{1}{|\mathbf{x}-\mathbf{y}|}$ is the fundamental solution of the Laplace equation. Physically, $\mathcal{V}$ maps an input electric charge distribution $\phi$ to the resulting electric potential distribution.

*Double layer potential.* The double layer potential $\mathcal{K} : H^{1/2}(\Gamma) \to H^{1/2}(\Gamma)$ is defined via

$$[\mathcal{K}\phi](\mathbf{x}) := \int_\Gamma \frac{\partial G(\mathbf{x}, \mathbf{y})}{\partial n(\mathbf{y})} \phi(\mathbf{y}) \, d\Gamma(\mathbf{y}),$$

where the integral is understood in the sense of the Cauchy principal value [Kanwal 2013]. Physically, $\mathcal{K}$ maps an input electric dipole density distribution $\phi$ to the resulting electric potential distribution.

*Adjoint double layer potential.* The adjoint double layer potential $\mathcal{T} : H^{-1/2}(\Gamma) \to H^{-1/2}(\Gamma)$ is defined as the conormal derivative of $\mathcal{V}$:

$$[\mathcal{T}\phi](\mathbf{x}) := \int_{\Gamma} \frac{\partial G(\mathbf{x}, \mathbf{y})}{\partial n(\mathbf{x})} \phi(\mathbf{y}) \, d\Gamma(\mathbf{y}),$$

where the integral is understood in the sense of the Cauchy principal value. Physically, $\mathcal{T}$ maps an input electric charge density distribution $\phi$ to the normal derivatives of the resulting electric potential distribution.

*Hypersingular operator.* The hypersingular operator $\mathcal{H} : H^{1/2}(\Gamma) \to H^{-1/2}(\Gamma)$ is defined as minus the conormal derivative of $\mathcal{K}$:

$$(\mathcal{H}\phi)(\mathbf{x}) := -\int_{\Gamma} \frac{\partial^2 G(\mathbf{x}, \mathbf{y})}{\partial n(\mathbf{x}) \partial n(\mathbf{y})} \phi(\mathbf{y}) \, d\Gamma(\mathbf{y}).$$

Physically, $\mathcal{H}$ maps an input electric dipole density distribution $\phi$ to normal derivatives of the resulting electric potential distribution.

*Operator properties.* $\mathcal{V}$ and $\mathcal{H}$ are self-adjoint operators, i.e.

$$\langle v, \mathcal{V}u \rangle_{\Gamma} \equiv \langle \mathcal{V}v, u \rangle_{\Gamma} \quad \text{and} \quad \langle v, \mathcal{H}u \rangle_{\Gamma} \equiv \langle \mathcal{H}v, u \rangle_{\Gamma},$$

where the inner product $\langle \cdot, \cdot \rangle_{\Gamma}$ is given by

$$\langle \phi, \psi \rangle_{\Gamma} := \int_{\Gamma} \phi(\mathbf{y})\psi(\mathbf{y}) \, d\Gamma(\mathbf{y}).$$

$\mathcal{K}$, $\mathcal{T}$ are adjoint to each other, i.e. $\langle v, \mathcal{K}u \rangle_{\Gamma} \equiv \langle \mathcal{T}v, u \rangle_{\Gamma}$. Furthermore, $\mathcal{V}$ is positive definite, and $\mathcal{H}$ is positive semidefinite.

### 3.3 Calderón projection & Dirichlet-to-Neumann operator

The Calderón projection relates $g$ and $g_n$ through the relationship

$$\begin{pmatrix} g \\ g_n \end{pmatrix} = \begin{pmatrix} \frac{1}{2}\mathcal{I} - \mathcal{K} & \mathcal{V} \\ \mathcal{H} & \frac{1}{2}\mathcal{I} + \mathcal{T} \end{pmatrix} \begin{pmatrix} g \\ g_n \end{pmatrix}$$

where $\mathcal{I}$ is the identity operator [Grubb 2009]. From the first row of this expression, we can derive the Dirichlet-to-Neumann operator $\mathcal{S} : g \mapsto g_n$ as the composition

$$\mathcal{S} = \mathcal{V}^{-1}\left(\frac{1}{2}\mathcal{I} + \mathcal{K}\right). \tag{6}$$

Combining both rows, however, reveals an alternative symmetric expression for the same operator:

$$\mathcal{S} = \mathcal{H} + \left(\frac{1}{2}\mathcal{I} + \mathcal{T}\right)\mathcal{V}^{-1}\left(\frac{1}{2}\mathcal{I} + \mathcal{K}\right). \tag{7}$$

Since $\mathcal{H}$ is self-adjoint and positive semidefinite, $\mathcal{V}$ is positive definite, and $(\mathcal{K}, \mathcal{T})$ are an adjoint pair, this alternative form symbolically shows that $\mathcal{S}$ is self-adjoint and positive semidefinite, with $\langle u, \mathcal{S}u \rangle_{\Gamma} = 0$ if and only if $u$ is a constant function. In our discretization, we work with this second formula because its discretization will become symmetric and positive semidefinite by construction. Note that interior Steklov eigenfunctions are known in closed form by the representation formula

$$u(\mathbf{x}) = \int_{\Gamma} \left[ G(\mathbf{x}, \mathbf{y})g_n(\mathbf{y}) - \frac{\partial G(\mathbf{x}, \mathbf{y})}{\partial n(\mathbf{y})} g(\mathbf{y}) \right] d\Gamma(\mathbf{y})$$

for any $\mathbf{x} \in \Omega$. Here, $u(\cdot)$ is the $i^{\text{th}}$ eigenfunction in the interior when we let $g(\cdot)$ and $g_n(\cdot)$ to be the $i^{\text{th}}$ eigenfunction on the surface and its normal derivative, respectively; later in equation (15) they are denoted as $\mathbf{u}$ and $\mathbf{t}$. Our algorithm will compute both $\mathbf{u}$ and $\mathbf{t}$, so interior eigenfunctions can be easily evaluated if needed.

## 4 DISCRETE DIRICHLET-TO-NEUMANN OPERATOR

### 4.1 Weak form boundary operators

We discretize all operators discussed above. Each corresponds to a weak-form discretized operator matrix:

$$\mathcal{V} \mapsto \mathbf{V}, \quad \mathcal{K} \mapsto \mathbf{K}, \quad \mathcal{T} \mapsto \mathbf{T}, \quad \mathcal{H} \mapsto \mathbf{H}, \quad \text{and} \quad \mathcal{I} \mapsto \mathbf{M}.$$

Take the single layer integral as an example. Assume $u(\Gamma)$ is a solution to the single layer integral equation $\mathcal{V}u(\mathbf{x}) = f(\mathbf{x})$. Then, for any test function $v : \Gamma \to \mathbb{R}$ we have the weak form $\langle v, \mathcal{V}u \rangle_{\Gamma} = \langle v, f \rangle_{\Gamma}$. Restricting $u, v, f$ to the piecewise-linear subspace $S_h^1(\Gamma)$ (the "hat functions" on a triangulated surface) leads to the finite-dimensional linear system $\mathbf{V}\mathbf{u} = \mathbf{M}\mathbf{f}$, where $\mathbf{V}, \mathbf{M}$ are Galerkin matrices

$$\mathbf{V} \in \mathbb{R}^{n \times n} : (\mathbf{V})_{ij} = \langle \phi_i, \mathcal{V}\phi_j \rangle_{\Gamma} \tag{8}$$

$$\mathbf{M} \in \mathbb{R}^{n \times n} : (\mathbf{M})_{ij} = \langle \phi_i, \phi_j \rangle_{\Gamma}. \tag{9}$$

In this linear system, $\mathbf{u}, \mathbf{f}$ contain the coefficients of $u$ and $f$ in the piecewise-linear basis, resp.

Similarly, we have

$$\mathbf{K} \in \mathbb{R}^{n \times n} : (\mathbf{K})_{ij} = \langle \phi_i, \mathcal{K}\phi_j \rangle_{\Gamma} \tag{10}$$

$$\mathbf{T} \in \mathbb{R}^{n \times n} : (\mathbf{T})_{ij} = \langle \phi_i, \mathcal{T}\phi_j \rangle_{\Gamma} \tag{11}$$

$$\mathbf{H} \in \mathbb{R}^{n \times n} : (\mathbf{H})_{ij} = \langle \phi_i, \mathcal{H}\phi_j \rangle_{\Gamma}. \tag{12}$$

It follows directly from properties in §3.2 that $\mathbf{V} = \mathbf{V}^{\mathsf{T}}$, $\mathbf{H} = \mathbf{H}^{\mathsf{T}}$, $\mathbf{T} = \mathbf{K}^{\mathsf{T}}$, $\mathbf{V} > 0$, and $\mathbf{H} \geq 0$. We refer interested readers to [Steinbach 2007, §2,3,4,5,6,7,10,12] for detailed discussion of these discretized operators.

Putting definitions together yields:

$$\mathbf{V}_{ij} = \sum_{\substack{T_1 \in A(i) \\ T_2 \in A(j)}} \iint_{T_1 \times T_2} \frac{1}{4\pi} \frac{1}{|\mathbf{x} - \mathbf{y}|} \phi_i(\mathbf{x})\phi_j(\mathbf{y}) \, d\Gamma_1(\mathbf{x}) \, d\Gamma_2(\mathbf{y}),$$

$$\mathbf{M}_{ij} = \sum_{T_1 \in A(i)} \int_{T_1} \phi_i(\mathbf{x})\phi_j(\mathbf{x}) \, d\Gamma_1(\mathbf{x}).$$

where $A(i)$ denotes the set of triangles adjacent to vertex $i$, and $\phi_i$ denotes the piecewise-linear "hat" basis function centered at vertex $i$. Note $\mathbf{M}_{ij}$ is same as the mass matrix used in FEM. To simplify notation for the remaining operators, define the following generic boundary operator $\mathbf{P}[p(\cdot, \cdot)] \in \mathbb{R}^{n \times n}$ w.r.t. kernel $p(\mathbf{x}, \mathbf{y})$:

$$\mathbf{P}_{ij}[p(\cdot, \cdot)] = \sum_{\substack{T_1 \in A(i) \\ T_2 \in A(j)}} \iint_{T_1 \times T_2} \frac{1}{4\pi} p(\mathbf{x}, \mathbf{y})\phi_i(\mathbf{x})\phi_j(\mathbf{y}) \, d\Gamma_1(\mathbf{x}) \, d\Gamma_2(\mathbf{y}). \tag{13}$$

Then, the discretized operators can be expressed as:

$$\mathbf{V}_{ij} = \mathbf{P}_{ij}[v], \quad \mathbf{K}_{ij} = \mathbf{P}_{ij}[k], \quad \mathbf{T}_{ij} = \mathbf{P}_{ij}[t], \quad \mathbf{H}_{ij} = \mathbf{P}_{ij}[h].$$

where

$$v(\mathbf{x}, \mathbf{y}) := \frac{1}{|\mathbf{x} - \mathbf{y}|},$$

$$k(\mathbf{x}, \mathbf{y}) := \frac{(\mathbf{x} - \mathbf{y}) \cdot n(\mathbf{y})}{|\mathbf{x} - \mathbf{y}|^3}, \quad t(\mathbf{x}, \mathbf{y}) := \frac{(\mathbf{y} - \mathbf{x}) \cdot n(\mathbf{x})}{|\mathbf{x} - \mathbf{y}|^3} = k(\mathbf{y}, \mathbf{x}),$$

$$h(\mathbf{x}, \mathbf{y}) := -\frac{n(\mathbf{x}) \cdot n(\mathbf{y})}{|\mathbf{x} - \mathbf{y}|^3} - \frac{3\left[(\mathbf{x}-\mathbf{y}) \cdot n(\mathbf{y})\right]\left[(\mathbf{y}-\mathbf{x}) \cdot n(\mathbf{x})\right]}{|\mathbf{x} - \mathbf{y}|^5},$$

and $n(\mathbf{x}), n(\mathbf{y})$ denote the normal directions at points $\mathbf{x}, \mathbf{y}$ on the surface, resp.

The matrix entries are evaluated using the Gaussian quadrature method implemented in [Šmigaj et al. 2015], which ensures quadrature points never coincide, even if the triangle $T_1$ is adjacent to or same as the triangle $T_2$ (these can happen, e.g., when evaluating diagonal entries that $i = j$).

### 4.2 Discretized Dirichlet-to-Neumann operator

We approximate the DtN operator by substituting the continuous operators with corresponding strong-form operators:

$$\mathcal{V} \to \mathbf{M}^{-1}\mathbf{V}, \quad \mathcal{K} \to \mathbf{M}^{-1}\mathbf{K}, \quad \mathcal{T} \to \mathbf{M}^{-1}\mathbf{T}, \quad \text{and} \quad \mathcal{I} \to \mathbf{I},$$

which leads to

$$\mathcal{S} \rightsquigarrow \mathbf{M}^{-1}[\mathbf{H} + (0.5\mathbf{M} + \mathbf{T})\mathbf{V}^{-1}(0.5\mathbf{M} + \mathbf{K})].$$

Define the (symmetric) weak form of this operator as

$$\mathbf{S} := \mathbf{H} + (0.5\mathbf{M} + \mathbf{T})\mathbf{V}^{-1}(0.5\mathbf{M} + \mathbf{K}). \tag{14}$$

## 5 MATRIX-FREE FORMULATION

Many applications require solving the linear system $\mathbf{Su} = \mathbf{Mf}$ or the generalized eigenvalue problem $\mathbf{Sx} = \lambda\mathbf{Mx}$. A naïve approach is to assemble all the operators as dense matrices via (14). Without acceleration, however, assembling these dense matrices would take $O(n^3)$ time, and matrix-vector products would require $O(n^2)$ operations. Furthermore, formula (14) involves inverting the dense matrix $\mathbf{V}$, which is expensive. Instead we apply a reformulation that avoids explicit matrix inversion.

### 5.1 Expansion

Define an auxiliary variable $\mathbf{t}$ as $\mathbf{t} := \mathbf{V}^{-1}(0.5\mathbf{M} + \mathbf{K})\mathbf{u}$. Notice that $\mathbf{t}$ represents the Neumann data, i.e. the normal derivative at boundary. The linear system $\mathbf{Su} = \mathbf{Mf}$ then can be expanded as a saddle point system

$$\begin{bmatrix} \mathbf{V} & -\mathbf{Q} \\ \mathbf{Q}^\mathsf{T} & \mathbf{H} \end{bmatrix} \begin{bmatrix} \mathbf{t} \\ \mathbf{u} \end{bmatrix} = \begin{bmatrix} 0 \\ \mathbf{Mf} \end{bmatrix},$$

where $\mathbf{Q} := 0.5\mathbf{M} + \mathbf{K}$.

The same technique also applies to reformulating the eigenvalue problem $\mathbf{Su} = \lambda\mathbf{Mu}$ as the system

$$\begin{bmatrix} \mathbf{V} & -\mathbf{Q} \\ \mathbf{Q}^\mathsf{T} & \mathbf{H} \end{bmatrix} \begin{bmatrix} \mathbf{t} \\ \mathbf{u} \end{bmatrix} = \lambda \begin{bmatrix} 0 & \\ & \mathbf{M} \end{bmatrix} \begin{bmatrix} \mathbf{t} \\ \mathbf{u} \end{bmatrix}. \tag{15}$$

The left-hand sides of these expressions do not contain matrix inverses, allowing us to use iterative linear system/eigenvalue solvers that only require matrix-vector products. See §5.4 for details of our solver and §5.6 for how we use the hierarchical techniques to *apply* $\mathbf{V}, \mathbf{Q}$, and other matrices without storing their elements.

### 5.2 Symmetrization

Numerical solution to saddle point systems like the ones in the previous subsection is a basic task in numerical PDE; see e.g. [Benzi et al. 2005]. To use the conjugate gradient (CG) algorithm and other techniques requiring positive definiteness, we apply a Bramble–Pasciak transformation [1988]:

$$\begin{bmatrix} \alpha\mathbf{V}\mathbf{P}_\mathbf{V}^{-1} - \mathbf{I} & 0 \\ -\alpha\mathbf{Q}^\mathsf{T}\mathbf{P}_\mathbf{V}^{-1} & \mathbf{I} \end{bmatrix} \begin{bmatrix} \mathbf{V} & -\mathbf{Q} \\ \mathbf{Q}^\mathsf{T} & \mathbf{H} \end{bmatrix} \begin{bmatrix} \mathbf{t} \\ \mathbf{u} \end{bmatrix} = \begin{bmatrix} 0 \\ \mathbf{Mf} \end{bmatrix}.$$

where $\mathbf{P}_\mathbf{V}$ is a symmetric preconditioner for the single layer potential (our choice is defined in §5.3). The constant $\alpha = 1/\gamma$ is chosen to ensure the positive-definiteness of the system matrix as $\gamma = 0.9\sigma_{\min}(\mathbf{P}_\mathbf{V}^{-1}\mathbf{V}) \leq 0.9 < 1$, where $\sigma_{\min}(\cdot)$ denotes the smallest singular value. The matrix of this system is symmetric and positive semidefinite:

$$\mathbf{A} = \begin{bmatrix} \alpha\mathbf{V}\mathbf{P}_\mathbf{V}^{-1}\mathbf{V} - \mathbf{V} & (\mathbf{I} - \alpha\mathbf{V}\mathbf{P}_\mathbf{V}^{-1})\mathbf{Q} \\ \mathbf{Q}^\mathsf{T}(\mathbf{I} - \alpha\mathbf{P}_\mathbf{V}^{-1}\mathbf{V}) & \mathbf{H} + \alpha\mathbf{Q}^\mathsf{T}\mathbf{P}_\mathbf{V}^{-1}\mathbf{Q} \end{bmatrix} = \mathbf{A}^\mathsf{T}. \tag{16}$$

Under the Bramble–Pasciak transformation, the eigenvalue problem can be written as

$$\mathbf{A} \begin{bmatrix} \mathbf{t} \\ \mathbf{u} \end{bmatrix} = \lambda \begin{bmatrix} 0 & \\ & \mathbf{M} \end{bmatrix} \begin{bmatrix} \mathbf{t} \\ \mathbf{u} \end{bmatrix}, \text{ or more simply } \mathbf{Ax} = \lambda\tilde{\mathbf{M}}\mathbf{x}.$$

Although $\mathbf{A}$ and $\mathbf{M}$ are positive semidefinite, $\alpha\mathbf{A} + \beta\tilde{\mathbf{M}}$ is strictly positive definite, for any $\alpha, \beta > 0$. So it can be solved by a shifted generalized eigenvalue method.

### 5.3 Preconditioning

Steinbach and Wendland [1998] prove that a modified hypersingular operator is a good preconditioner for the single layer potential and that the single layer potential is a good preconditioner for the Dirichlet-to-Neumann operator (14). In particular, they show that the spectral condition numbers of these two preconditioners are $O(1)$, in the sense that the condition number remains the same when the domain is upsampled. Their choices of preconditioners for $\mathbf{S}$ and $\mathbf{V}$ are:

$$\mathbf{P}_\mathbf{V}^{-1} := 4\mathbf{M}^{-1}\left(\mathbf{H} + \frac{\beta}{4}\mathbf{1}\mathbf{1}^\mathsf{T}\right)\mathbf{M}^{-1} \tag{17}$$

$$\mathbf{P}_\mathbf{S}^{-1} := \mathbf{M}^{-1}\mathbf{V}\mathbf{M}^{-1}$$

where $\beta = 1/(\mathbf{1}^\mathsf{T}\mathbf{M}\mathbf{1})^{3/2}$ is chosen to ensure scale invariance. Since $\mathbf{M}$ is sparse and diagonally dominant, Jacobi iteration could compute $\mathbf{M}^{-1}\mathbf{x}$ in $O(n)$ time. In our implementation, however, we use a diagonal lumped mass matrix that can be inverted in closed form.

For saddle point system $\mathbf{Ax} = \lambda\tilde{\mathbf{M}}\mathbf{x}$, Bramble and Pasciak [1988] recommend the following preconditioner to (16), which is spectrally equivalent to $\mathbf{A}$:

$$\mathbf{P}_\mathbf{A}^{(1)} := \begin{bmatrix} \mathbf{V} - \gamma\mathbf{P}_\mathbf{V} & \\ & \mathbf{P}_\mathbf{S} \end{bmatrix}.$$

The upper left block of this matrix cannot be inverted efficiently, so it has to be combined with a modified Bramble-Pasciak CG solver. See [Stoll and Wathen 2007] for a survey on other preconditioning options for saddle point system. In practice, we use the following simpler alternative that is easier to invert:

$$\mathbf{P}_\mathbf{A}^{(2)} := \begin{bmatrix} \frac{1}{\alpha-1}\mathbf{P}_\mathbf{V} & \\ & \mathbf{P}_\mathbf{S} \end{bmatrix}.$$

$\mathbf{P}_A^{(2)}$ is a valid preconditioner as well, that is, $\mathbf{P}_A^{(2)}$ is spectrally equivalent to $\mathbf{A}$, since $\mathbf{P}_A^{(2)}$ is spectrally equivalent to $\mathbf{P}_A^{(1)}$ and $\mathbf{P}_A^{(1)}$ is spectrally equivalent to $\mathbf{A}$.[1] We find this choice suffices for our solver; investigating the most efficient preconditioning strategy is a potential topic for future work.

### 5.4 Iterative solvers

We use the preconditioned conjugate gradient (CG) algorithm to solve linear systems of equations involving our positive definite operators. As noted above, PCG requires matrix-vector products rather than storing our operators explicitly, as would be required by Gaussian elimination.

We employ the locally optimal block preconditioned conjugate gradient (LOBPCG) solver [Knyazev 2001] to compute the spectrum efficiently. The convergence speed of this method depends on the condition number $\text{cond}(\mathbf{P}^{-1}\mathbf{A})$. LOBPCG allows for an initial guess of the eigenvectors. We apply a multi-scale approach, computing "progressive spectra" from progressively simplified meshes obtained using [Hoppe 1996]. We first compute eigenvalues and eigenfunctions on a low-resolution mesh and then upsample the low-resolution eigenfunctions as initialization for high-resolution meshes. Typically we use 2–3 levels of progressively simplified meshes, where at each level the mesh is simplified by a factor of $1/4$; we use the nearest neighborhood rule to upsample eigenfunctions. Note Vaxman et al. [2010] use a related approach to evaluate the heat kernel.

### 5.5 Generalization

We have introduced the Dirichlet-to-Neumann operator for compact manifolds with closed boundaries. As long as we are given a normal vector field at every point, however, the integral-based definitions of inner products $\langle v, \mathcal{V}u \rangle_\Gamma$, $\langle v, \mathcal{K}u \rangle_\Gamma$, $\langle v, \mathcal{H}u \rangle_\Gamma$, and $\langle v, \mathcal{T}u \rangle_\Gamma$ as well as the resulting operator $\mathcal{S}$ remain valid even if $\Gamma$ does not bound some region in $\mathbb{R}^3$. Given this observation, to process oriented surfaces with boundary we simply use the same DtN operator (14), although without theoretical justification via PDE.

We experimentally verify that this natural generalization demonstrates reasonable and robust behavior. Figure 2 illustrates our "generalized Steklov" eigenfunctions computed on a hemisphere from these integral operators. Figure 3 demonstrates that the resulting spectrum blends smoothly as the shape changes from a sphere to a hemisphere by moving the cutting plane linearly.

### 5.6 Implementation details

Our rephrasing of the DtN operator in terms of boundary integrals is an example of the boundary element method (BEM) from

$$\sigma_{\min}(\mathbf{P}_A^{(1)^{-1}}\mathbf{A})\langle \mathbf{x}, \mathbf{x} \rangle_{\mathbf{P}_A^{(1)}} \leq \langle \mathbf{x}, \mathbf{x} \rangle_A \leq \sigma_{\max}(\mathbf{P}_A^{(1)^{-1}}\mathbf{A})\langle \mathbf{x}, \mathbf{x} \rangle_{\mathbf{P}_A^{(1)}}$$

$$\sigma_{\min}(\mathbf{P}_V^{-1}\mathbf{V})\langle \mathbf{x}, \mathbf{x} \rangle_{\mathbf{P}_V} \leq \langle \mathbf{x}, \mathbf{x} \rangle_V \leq \sigma_{\max}(\mathbf{P}_V^{-1}\mathbf{V})\langle \mathbf{x}, \mathbf{x} \rangle_{\mathbf{P}_V}$$

where $\langle \mathbf{x}, \mathbf{x} \rangle_B := \mathbf{x}^\top\mathbf{Bx}, \forall\mathbf{B}$. These lead to $\sigma_1\langle \mathbf{x}, \mathbf{x} \rangle_{\mathbf{P}_A^{(2)}} \leq \langle \mathbf{x}, \mathbf{x} \rangle_A \leq \sigma_2\langle \mathbf{x}, \mathbf{x} \rangle_A$, where $\sigma_1 := \min\left\{1, \frac{1}{9}\left(1-\gamma\right)\right\}$ and $\sigma_2 := \max\left\{1, (1-\gamma)\left\lceil\frac{\sigma_{\max}(\mathbf{P}_V^{-1}\mathbf{V})}{\gamma}-1\right\rceil\right\}$.

---

[1] More precisely, by the definition of preconditioners we have:

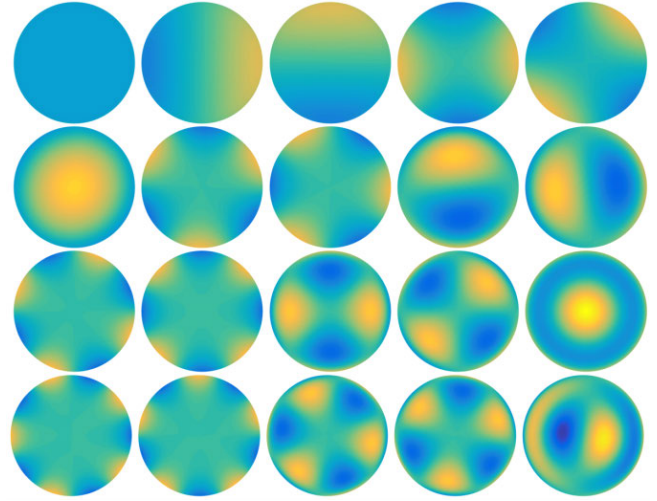

Fig. 2. Steklov eigenfunctions of a hemisphere.

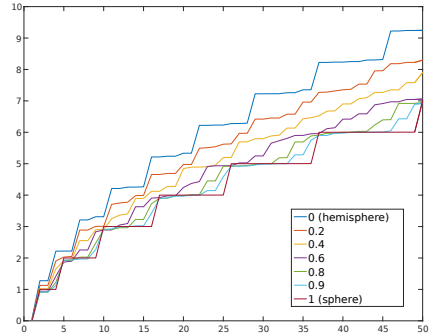

Fig. 3. Generalized Steklov eigenvalues of a sphere cut with a moving plane.

numerical PDE. Hence, our implementation uses the BEM++ library [Śmigaj et al. 2015] to evaluate the matrices above, whose elements are obtained using numerical quadrature. BEM++ uses hierarchical matrices ("H-matrices") to compute matrix-vector products efficiently, approximating the integrals with hierarchical expressions that group together small contributions from far-away vertices [Börm et al. 2003]. $\sigma_{\min}(\mathbf{P}_V^{-1}\mathbf{V})$ can be approximated by solving the minimum eigenvalue of the squared system, without the need for a preconditioner.

## 6 VALIDATION

### 6.1 Robustness to topological change

In addition to the hemisphere test in Figure 2, Figure 4 gives an example of robustness of our operator to topological change. Here, we compare the Steklov eigenvalues and eigenfunctions of a torus with zero, one, and two rings cut out. When triangles on the cut are removed, these topological changes—which drastically affect the Laplace–Beltrami operator—have little effect on the Steklov spectrum. When we close the holes with flat disk patches, however, the Steklov spectrum detects a topological change, since the donut is clearly divided into two volumetric pieces.

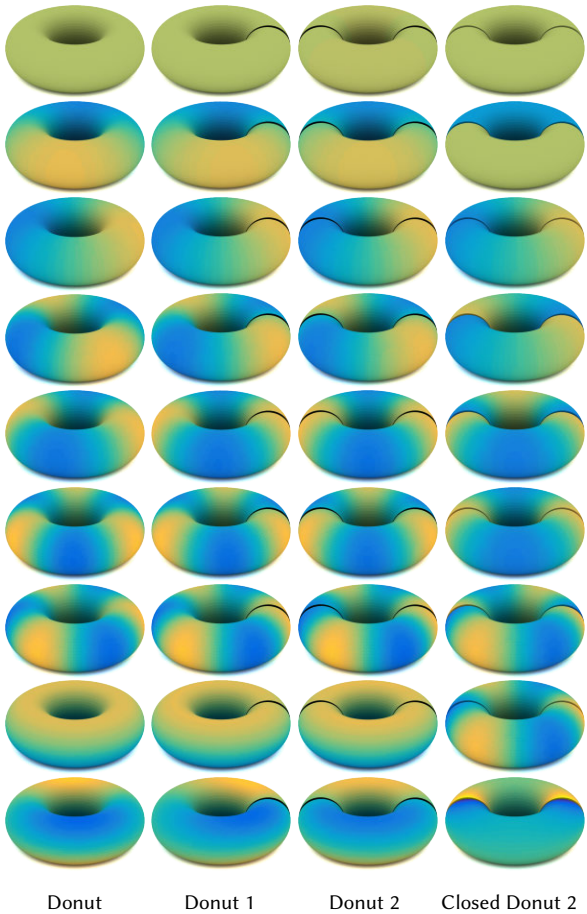

Donut   Donut 1   Donut 2   Closed Donut 2

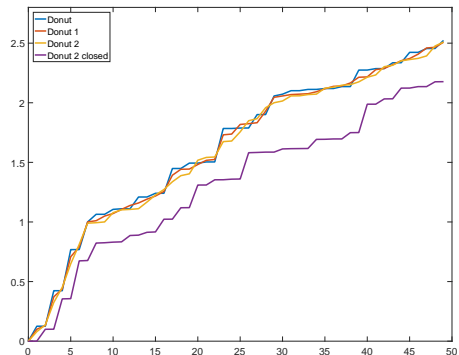

Fig. 4. The Steklov spectra of the donuts. These measurements remain stable as the donut undergoes topological change and even becomes disconnected. In the last column, the two open holes on the cut donut are closed with disks, in which case the Steklov spectrum considers the donut as two disconnected pieces.

Figure 5 shows an example of robustness to "topological noise." Here, we remove triangles from the mesh with uniform probability and show the resulting Steklov spectrum. Once again, our operator is remarkably stable to these changes, yielding a stable spectrum even when 50% of the triangles are removed.

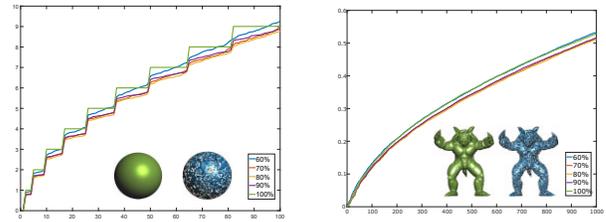

Fig. 5. The Steklov spectrum is robust when removing random triangles progressively from the mesh, tested on the sphere and armadillo meshes.

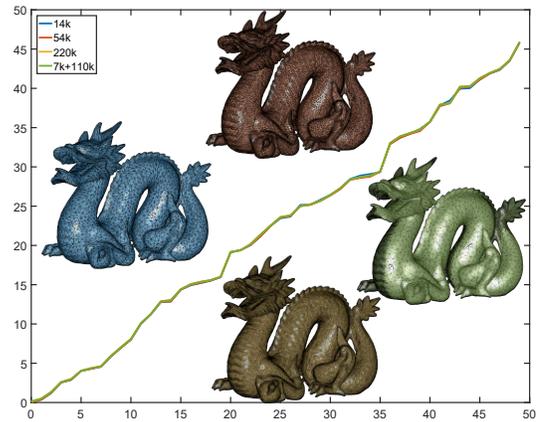

Fig. 6. The Steklov spectrum is robust to downsampling as well as unbalanced sampling. Dragons represented with 14k (blue), 54k (red), 220k (yellow), and 117k (green) triangles have very similar Steklov eigenvalues. The dragon with 117k triangles are obtained by downsampling only the left side of the 220k dragon.

## 6.2 Robustness to surface sampling

Figure 6 shows an example of robustness to downsampling and unbalanced sampling. Here we simplify the dense dragon mesh and observe that the resulting Steklov spectrum does not change significantly. Although geometric details at the belly and scales of the dragon are lost during downsampling, the eigenvalues are almost unaffected.

Since the low-order Steklov eigenvalues are robust to downsampling and in most applications we are interested in only the first ∼50 eigenfunctions, the experiment suggests it is not necessary to use a highly-detailed mesh. We find a mesh with $7k-15k$ triangles is usually sufficient to accurately approximate the top 50 eigenfunctions; in this regime, computation finishes in $4-10$ minutes.

## 6.3 Robustness to vertex noise

Figure 7 illustrates stability of our BEM discretization when noise is added to the boundary surface. Although they are defined in §3.1 via a second-order differential equation, we observed that the lower Steklov eigenfunctions are particularly robust to noise, similarly to those of the Laplace–Beltrami operator. Both Laplacian and Steklov spectra scale down globally, since adding noise increases the edge lengths and thus the surface area of the mesh.

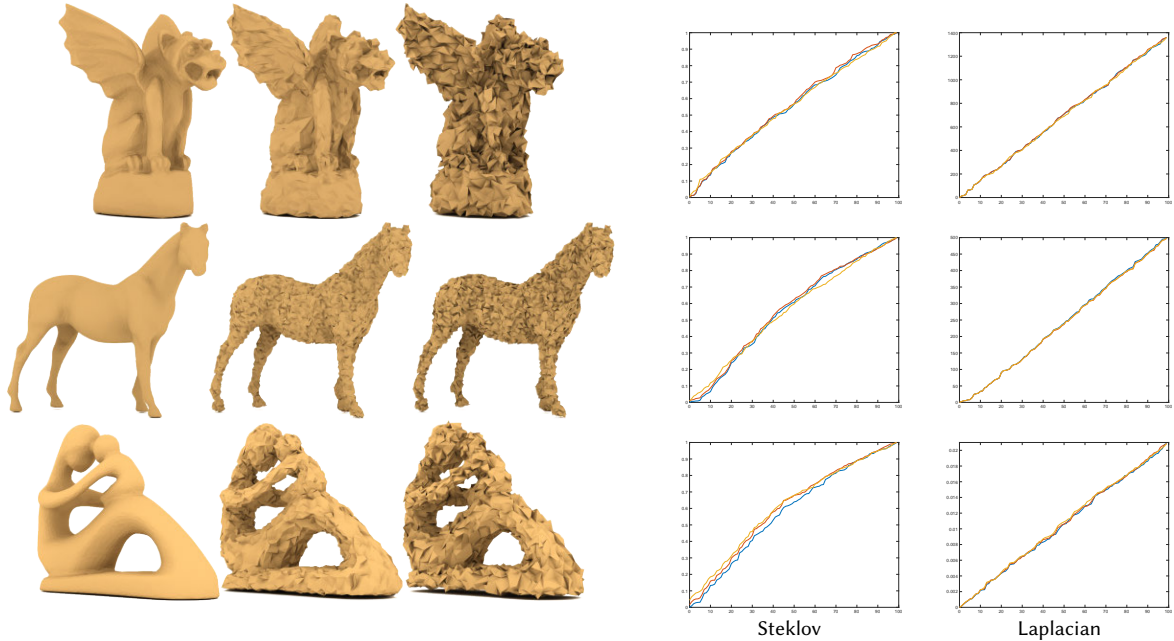

Fig. 7. Robustness test by adding noise to the vertex positions. Blue, red and yellow curves corresponds to the original and noisy meshes, respectively. After noise is applied, the mesh may not be exactly watertight due to self-intersections as well as many flipped triangles, and hence the smallest "Steklov eigenvalue" could be slightly positive. A decrease of eigenvalues after adding noise is expected since surface area increases. Both eigenvalues are normalized by its largest eigenvalue.

When noise is applied, the mesh is no longer watertight: There are intersecting triangles, flipped triangles, and so on caused by random vertex perturbations. For these challenging meshes, the BEM++ library—not designed to handle these cases—occasionally fails because the quadrature points can coincide.

To avoid this issue, we first apply a simple repair procedure to the mesh: Cut intersecting triangles into multiple non-intersecting ones, merge vertices that are closer than $\delta$ to each other, and remove duplicated triangles (if there are any). $\delta = 10^{-2}\bar{L}$ is chosen, where $\bar{L}$ is the average edge length in the original input mesh. This repair procedure leads to almost no visual difference to the mesh. In addition, for these extremely noisy tests we find using a regularized kernel $1/(r + \epsilon)$ instead of the original kernel $1/r$ helps to improve solver's robustness, where $\epsilon = 10^{-4}$ is chosen, assuming the mesh has been normalized to the unit scale. This simple repair procedure and regularization are only necessary for meshes with largest amount of noise in Fig. 7 (rightmost column), to avoid overlapping quadrature points.

Our Steklov solver remains robust even with a significant amount of noise. Large noise may cause partial occlusions (like the patched disks in Fig. 4), which the Steklov spectrum considers as partial topological change. In this case, the smallest computed eigenvalue could be slightly positive, depending on the size of the occlusion.

As a point of contrast, no volumetric (tet-based) method can be applied before fixing the mesh to be watertight; this more complex repair procedure is challenging and may lose geometric information. In a sense, it is not surprising that our BEM-based approach is robust to noise and triangle soup; the state-of-the-art mesh robust repair technique [Jacobson et al. 2013] in essence is also a boundary element method.

### 6.4 Stability test for volume isometries

The Steklov spectrum is invariant to volumetric isometry, and our experiments demonstrate that it remains stable when the volume enclosed by the outer surface deforms *near*-isometrically. In the first row of Figure 8, when the rectangular prisms are deformed using a near-volume-isometric bending map, the Steklov and Laplacian eigenvalues and eigenfunctions remain almost unchanged. In the second row, however, the thin box is deformed subject to a near-surface-isometry but non-volume-isometry, by shearing the interior of the box. Only the Steklov spectrum captures the volumetric change as the shear increases. This example illustrates how the Steklov spectrum discriminates nuanced non-isometric volume changes, while the Laplacian spectrum fails to capture the difference.

### 6.5 Conditioning

Convergence of our iterative linear and eigenvalue solvers depends on the conditioning of the matrices involved. Since our operators and preconditioners are constructed using integrals that make sense in the continuum, we expect that conditioning is $O(1)$ in mesh size, i.e., it does not depend significantly on the number of vertices or triangles. Figure 9 verifies this relationship for the single layer potential operator on different meshings of the unit sphere; we observe similar behavior on other meshes.

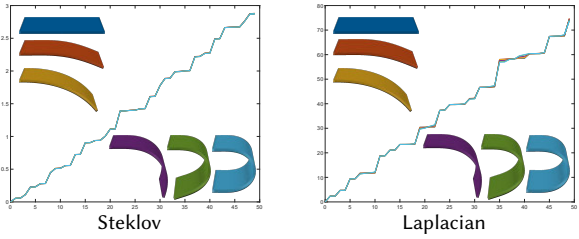

(a) Both Steklov and Laplacian spectra are robust to (near)-volume-isometry.

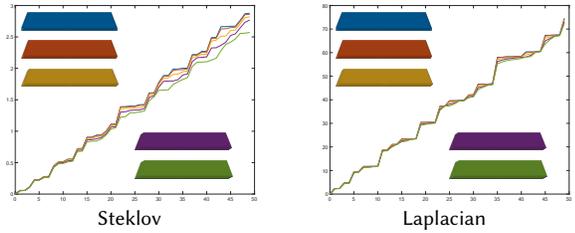

(b) Steklov spectrum captures shearing, to which Laplacian is insensitive.

Fig. 8. Stability tests for volume-isometries and non-volume-isometries.

| # vertices | # faces | cond($P_V^{-1}V$) |
|---|---|---|
| 258 | 512 | 3.91 |
| 1026 | 2048 | 3.96 |
| 4098 | 8192 | 4.00 |
| 16386 | 32768 | 4.00 |
| 65538 | 131072 | 3.94 |
| 262146 | 524288 | 3.95 |

Fig. 9. Conditioning on different meshings of the unit sphere.

### 6.6 Convergence to analytical eigenvalues

The analytical Steklov eigenvalues and eigenfunctions are known for the sphere. We compare the numerical solution of our method on the sphere for verification purposes. Figure 10 verifies that our discretized operators are faithful to these ground-truth quantities.

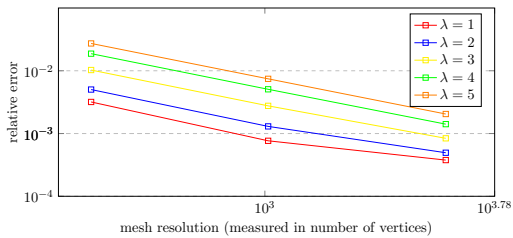

Fig. 10. Relative error measured in between spherical harmonics and the eigenspaces of our Steklov operator. Relative error decreases as the mesh resolution increases. The error is measure for a fixed number of 25 iterations, computed from random initialization.

### 6.7 Timing

For the cube with 6146 vertices and 12288 triangles, BEM++ takes 60 seconds to assemble all boundary element operators, one LOBPCG iteration takes 24 seconds on average, and 20 iterations are usually sufficient for random initialization. Usually less than 10 iterations are needed for initialization from low-resolution eigenfunctions. We observed that all the timings roughly scale linearly with size of the mesh. Our implementation is in PYTHON, and results are collected on an Alienware laptop with an Intel i7 4800MQ CPU and Linux operating system. As many graphics applications do not require a very high precision, we believe that there is large room for future work to improve the computational efficiency (§8).

### 6.8 Comparison with FEM

An alternative approach for extrinsic spectral geometry uses the volumetric Laplacian. Spectral computations with this operator do not have a boundary integral formulation, and hence we would need a volumetric mesh of the interior of the domain. This limits applicability, since tet meshing typically requires a watertight surface as input, and the results depend on the algorithm used for generating a tet mesh. Furthermore, map-based applications such as computation of shape differences (§7.5) would require correspondence of not just the outer surface but the interior as well, which may not be available. As we will show in §7.2, we do not observe a case in which an explicit mesh of the interior leads to significantly better results for geometry processing tasks.

These high-level concerns aside, one could reasonably discretize the DtN operator using the finite element method (FEM) on a tet mesh. In particular, the DtN operator can be derived as the Shur complement of the volumetric Laplacian matrix $L^{(vol)}$, i.e. FEM discretization of the interior Laplace equation with the Neumann boundary condition. With piecewisely linear bases, $L^{(vol)}$ is assembled by the familar cotangent weights (of the dihedral angle opposite to an edge); see e.g. [Jacobson 2013, §2.1] for an explicit formula . On a tetrahedral mesh with volumetric FEM-based Laplacian $L^{(vol)}$ and mass matrix $M^{(vol)}$, divide this matrix into blocks corresponding to boundary vertices $b$ and interior vertices $i$:

$$L^{(vol)} = \begin{bmatrix} L_{bb} & L_{bi} \\ L_{ib} & L_{ii} \end{bmatrix}.$$

Then, we discretize the Laplace equation $\Delta u = 0$ in the interior as

$$\begin{bmatrix} \nabla_{\mathbf{n}} \mathbf{u}_b \\ \Delta \mathbf{u}_i \end{bmatrix} = M^{(vol)-1} L^{(vol)} \mathbf{u}.$$

Setting the second row $\Delta \mathbf{u}_i = 0$ to eliminate the interior vertices as $\mathbf{u}_i = -L_{ii}^{-1} L_{ib} \mathbf{u}_b$, the Dirichlet energy takes the form

$$E = \mathbf{u}^\top L^{(vol)} \mathbf{u} = \mathbf{u}_b^\top S^{(FEM)} \mathbf{u}_b,$$

giving rise to a FEM-based Dirichlet-to-Neumann operator

$$S^{(FEM)} = L_{bb} - L_{bi} L_{ii}^{-1} L_{ib} \geq 0.$$

Figure 11 illustrates a drawback of using the tet-based DtN operator. Here, we compute Steklov eigenvalues on two different tetrahedral meshes of the same unit sphere using first-order FEM. This figure shows that Steklov eigenvalues depend on the choice of tetrahedral meshing of the boundary. In particular, faster sparse computations on coarse FEM meshes come at the cost of inaccurate Steklov approximation. Contrastingly, BEM faithfully approximates Steklov eigenvalues for the sphere beyond the range tested in this experiment.

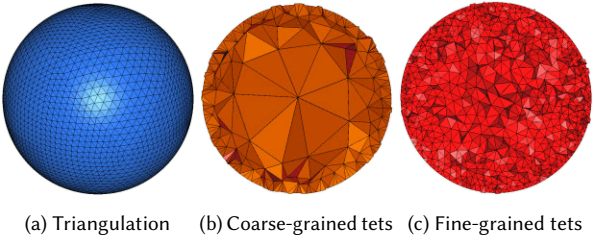

(a) Triangulation    (b) Coarse-grained tets    (c) Fine-grained tets

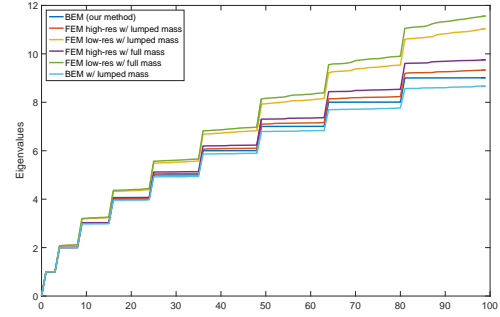

(d) Plot of eigenvalues computed using different methods.

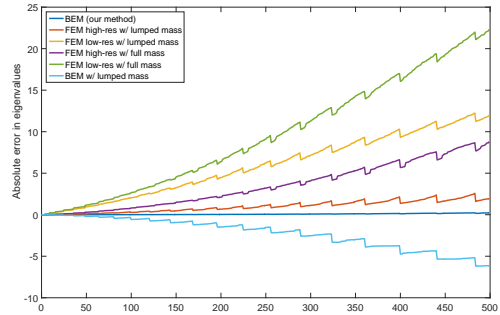

(e) Plot of absolute errors of eigenvalues (numerical solutions minus analytical solutions). For the unit sphere, the analytical solution of the Steklov eigenfunctions and eigenvalues are spherical harmonics and corresponding eigenvalues).

Fig. 11. Steklov spectrum of unit sphere, computed using BEM and FEM with different mesh resolutions. For FEM, we compute the Steklov spectrum using either fine-grained tets or coarse-grained tets, and using either a full mass matrix or a lumped mass matrix. For BEM, we experiment with full and lumped mass matrices.

We further observe that BEM with the correct full mass matrix (our method) yields the closest spectrum to ground truth by a significant margin; even BEM with a lumped diagonal mass matrix does not yield accurate results. Hence, our implementation always uses the full mass matrix in the generalized eigenvalue problem. We do use lumped mass matrices to construct preconditioners for the iterative solvers, since they can be inverted quickly; this affects only the number of iterations for the solver rather than the quality of the final output.

## 7 EXPERIMENTS AND APPLICATIONS

We highlight two properties of the DtN operator $\mathcal{S}$ distinguishing it from the Laplace–Beltrami operator, motivated in the theoretical discussion in §3:

- $\mathcal{S}$ captures the volumetric structure of the interior.
- $\mathcal{S}$ incorporates the mean curvature of the outer surface.

Accordingly, the algorithms we derive based on the Steklov spectrum enjoy these properties.

Many existing frameworks for intrinsic geometry processing can be extended to take extrinsic geometry into consideration by substituting the Laplacian operator with the DtN operator. There are many ways to justify this simple substitution:

- Many intrinsic geometry tasks involve the surface-based Laplace–Beltrami operator, whose weak form corresponds to a Dirichlet energy on the surface:

$$\langle \nabla_\Gamma u, \nabla_\Gamma u \rangle_\Gamma = \mathbf{u}^\mathsf{T} \mathbf{L} \mathbf{u}.$$

The weak form of the DtN operator provides the closely-related *volumetric* Dirichlet energy of the harmonic extension $\mathcal{E}$ of a surface-based function:

$$\langle \nabla \mathcal{E} u, \nabla \mathcal{E} u \rangle_\Omega = \mathbf{u}^\mathsf{T} \mathbf{S} \mathbf{u}.$$

- Both operators' associated eigenproblems involve Rayleigh quotients with similar numerators and identical denominators [Girouard and Polterovich 2017].
- For $n$ disconnected/closed pieces, both operators have $n$-dimensional null spaces.
- Laplace–Beltrami and DtN operators both can be understood as subtracting the value of a function at a point from its average over a neighborhood. The DtN operator first interpolates the function harmonically to the volume, while Laplace–Beltrami restricts to the surface.

The close analogies above hint that shifting from intrinsic geometry to extrinsic geometry can be accomplished by substituting the cotangent Laplacian operator with a discretized DtN operator.

We demonstrate this substitution in a few contexts. In particular, we experiment and study properties of Steklov-based kernel signatures, distances, and shape differences, which are fundamental components of many high-level geometry processing and shape analysis algorithms.

### 7.1 Steklov spectrum

Figures 12 and 13 illustrate Steklov eigenfunctions on more complex models; surface-based Laplace–Beltrami eigenfunctions are shown for comparison. Both models have more vertices than can be feasibly handled using dense matrix computations, highlighting the necessity for iterative solvers introduced in §5.4.

A few qualitative differences between the intrinsic and extrinsic eigenfunctions are worth pointing out. On the gargoyle model, the Steklov eigenfunctions exhibit more localization on the wings; the wings are distinctive volumetric features, whereas the outer surface blends the wings into the body/base. On the dragon model, eigenfunctions of the two operators look completely different: Steklov eigenfunctions are localized in individual folds of the dragon, while the Laplace–Beltrami eigenfunctions extend along the entire surface.

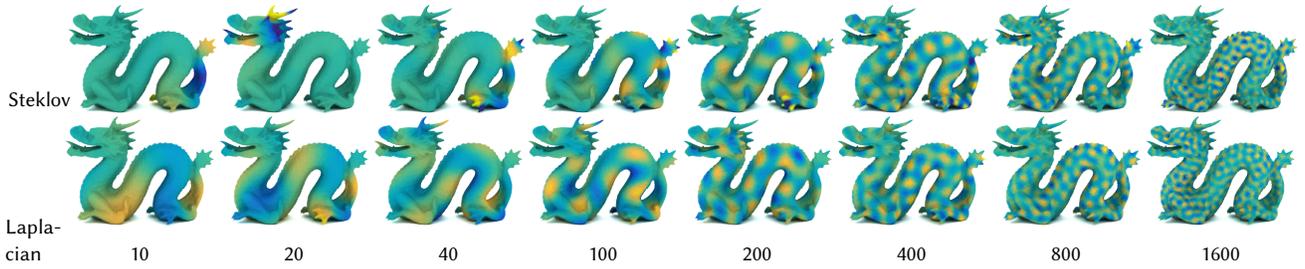

Fig. 12. Selected Steklov eigenfunctions on the dragon model, with comparison to Laplacian eigenfunctions. Although the Steklov eigenfunctions are computed using the boundary geometry, these eigenfunctions are aware of the geometry of the volume enclosed by the dragon surface, in particular the S-shaped bend of its body.

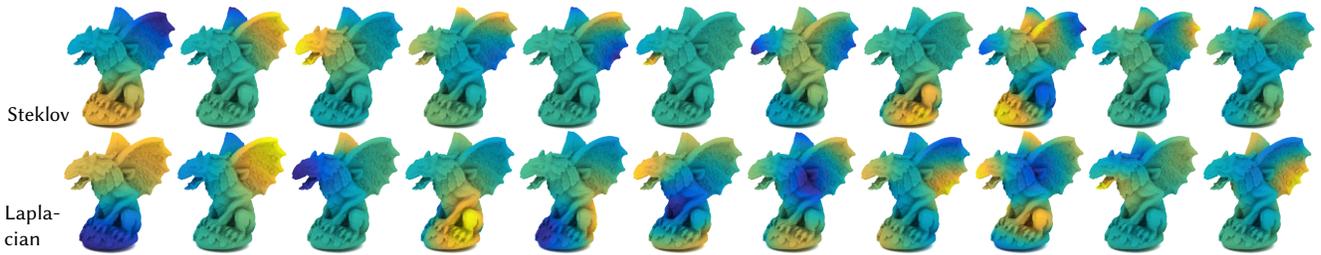

Fig. 13. The Steklov eigenfunctions corresponding to the smallest 12 eigenvalues, compared to the surface Laplacian eigenfunctions; this model contains 50k triangles and 25k vertices.

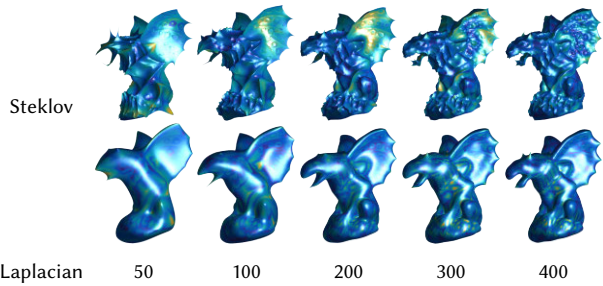

Fig. 14. Reconstruction with increasing number of bases; color encodes relative error in position.

This effect is reflective of the fact that intrinsically the dragon model is a long tube in which the 180° fold of the body between the front and hind legs is insignificant; volumetrically, however, the up/down bending of the dragon is a prominent geometric feature.

Figure 14 shows an experiment similar to the ones in [Vallet and Lévy 2008], in which the eigenfunctions of the DtN operator are used to compress and subsequently reconstruct the $xyz$ coordinate functions of a triangle mesh. Interestingly, fewer Steklov eigenfunctions are needed to capture key geometric features relative to the Laplace–Beltrami eigenfunctions. That is, low-frequencies in DtN space appear to better capture geometric variation.

### 7.2 Kernel-based descriptors

The heat equation associated to the DtN operator is given by

$$\frac{\mathrm{d}u}{\mathrm{d}t} = -\mathcal{S}u, \tag{18}$$

whose solution can be written

$$u(x,t) = \int_{\Gamma} k_t(x,y)f(y)\,\mathrm{d}y. \tag{19}$$

Here, $k_t(x,y)$ is the heat kernel

$$k_t(x,y) = \sum_{i=0}^{\infty} e^{-\lambda_i t}\phi_i(x)\phi_j(y), \tag{20}$$

where $\lambda_i$ and $\phi_i$ are the $i^{\text{th}}$ eigenvalue and eigenfunction of $\mathcal{S}$, resp.

If we replace DtN with the Laplace–Beltrami operator, the *heat kernel signature* (HKS) of a point $x \in \Gamma$ is defined as $h_t(x) = k_t(x,x)$ [Sun et al. 2009], the diagonal of the heat kernel. Considered as a function of $t$, $h_t(x)$ provides a multi-scale characterization of intrinsic geometry near $x$; it is a basic descriptor used in many shape matching and correspondence algorithms. The wave kernel signature (WKS) [Aubry et al. 2011], also defined using the Laplace–Beltrami operator, outperforms the heat kernel signature for shape matching tasks using a related eigenvalue formula.

The heat and wave kernel signatures can be naturally generalized to the DtN operator by replacing the Laplace–Beltrami spectrum/eigenfunctions with Steklov spectrum/eigenfunctions. A similar construction is considered by Raviv et al. [2010], who construct a volumetric HKS via a coarse discretization of the Laplacian in the interior of a volume bounded by a watertight surface. Without this interior meshing, our DtN-based HKS and WKS capture similar extrinsic shape properties.

Figure 15 compares the Laplace–Beltrami, DtN, and volumetric HKS functions on a triangulated surface; the volumetric HKS is approximated using a tetrahedral mesh of the interior, as explained in §6.8. Following Sun et al. [2009], we consider the time interval $[\frac{4\ln 10}{\lambda_{300}}, \frac{4\ln 10}{\lambda_2}]$ in all of our examples, and Figure 15 shows typical patterns of the signature using either a small or large time from the range. The DtN and volumetric kernel signatures capture the mean curvature in the palm of the hand, ignored by the intrinsic HKS. Similarly, Figure 16 illustrates the point signatures at the apexes of the bumps in Figure 1. Since the two cubes are isometric, Laplace–Beltrami-based signatures does not discriminate the two points, while the DtN-based signature distinguishes the two models.

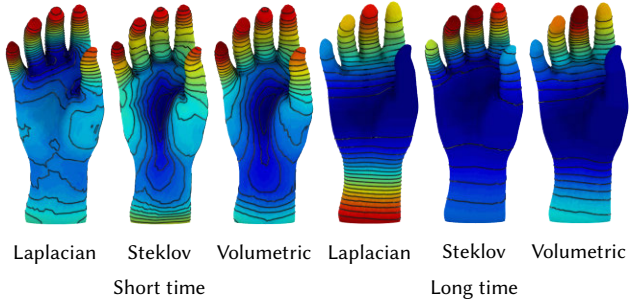

| Laplacian | Steklov | Volumetric | Laplacian | Steklov | Volumetric |
|---|---|---|---|---|---|

Fig. 15. Visualization of Laplacian and Steklov heat kernel function $h_t(x)$, respectively, in log scale. Note that the minima correspond to most negative Gaussian and mean curvatures, respectively.

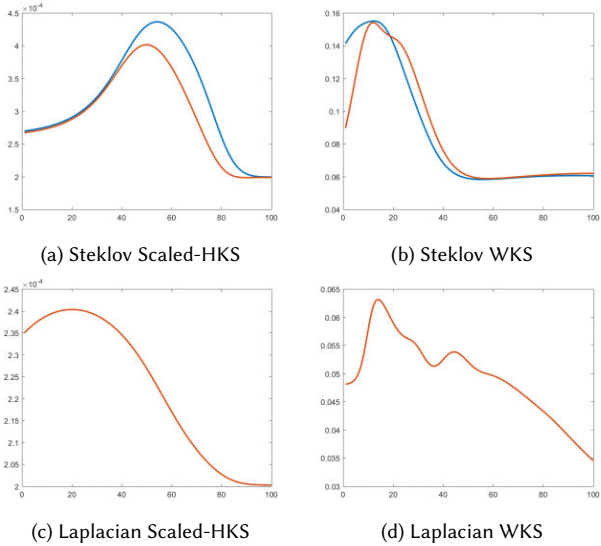

(a) Steklov Scaled-HKS     (b) Steklov WKS

(c) Laplacian Scaled-HKS     (d) Laplacian WKS

Fig. 16. Steklov-based scaled heat kernel signature and wave kernel signature on the cubes with inward (red) and outward (blue) bumps. In contrast, the two cubes have identical Laplacian-based signatures. Feature points are chosen as centers of the bumps.

Figure 17 illustrates how DtN-derived descriptors can be more discriminative than their Laplace–Beltrami counterparts. Here, we mark a point on the foot of a human model and show the distance between its descriptor and those of other surface points. Fewer points have descriptors close to the descriptor of the foot, showing that the DtN WKS embedding is less ambiguous. In the second example, DtN WKS is better aware of the left-right symmetry breaking, and tends to discriminate left and right feet. This property may or may not be desirable depending on application, but it shows that DtN operators are *pose-aware* and can be used to navigate databases of near-isometric models like articulated humans.

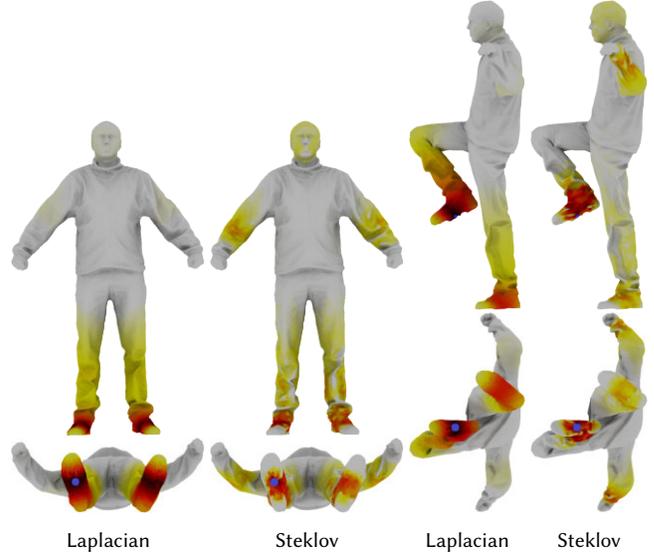

| Laplacian | Steklov | Laplacian | Steklov |
|---|---|---|---|

Fig. 17. Color encodes the similarity between the signatures across the shape with signature at the marker (blue dot), which sits in the center of shoe bottom. Darker color corresponds to more similarity. *Left*: Steklov WKS is aware of regions with a large mean curvature (wrinkles on the cloth) and avoids matching them with the marker, which is on a plateau with zero mean curvature. *Right*: Steklov WKS is a more restrictive signature, while Laplacian WKS tends to confuse left and right feet.

### 7.3 Spectral distance

Several spectral distances are defined between points on a surface in terms of the spectrum of the Laplace–Beltrami operator. These distances enjoy certain stability and smoothness properties not satisfied by the geodesic distance and are computable using linear algebra machinery. These distances naturally generalize to the Steklov spectrum, providing spectral volume-aware distances that require computation only on the boundary.

Working in analogy to previous work [Coifman and Lafon 2006; Lipman et al. 2010], we define the Steklov diffusion distance and the bi-Steklov distance as follows

$$d_D(x,y)^2 = \sum_{i=1}^{\infty} e^{-2t\lambda_i} \left( \phi_i(x) - \phi_i(y) \right)^2$$

$$d_B(x,y)^2 = \sum_{i=1}^{\infty} \frac{1}{\lambda_i^2} \left( \phi_i(x) - \phi_i(y) \right)^2 .$$

These distances are aware of relationships between points that reach across the volume.

Figures 18, 19, and 20 show several examples of these distances on meshed surfaces. A few key examples point out the special properties of our distance compared to its intrinsic and/or more naïve extrinsic counterparts:

• FLAT DISK (FIGURE 18): The DtN distance is small between points at the center of the top and the bottom of the disk, which are

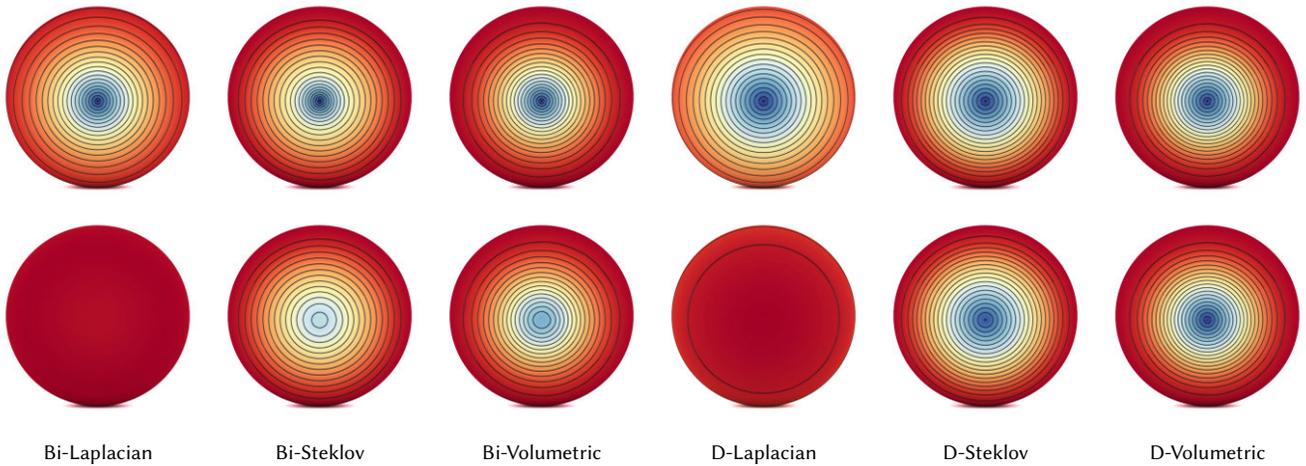

Fig. 18. Spectral and diffusion distances computed using the Laplacian, Steklov, and volumetric Laplacian spectra; colors range from blue (zero distance) to red (large distance). The first and second rows show the top and bottom of the surface, respectively. Source point is placed at the center of the top surface (visible in the first row). The Steklov (boundary-based) and volumetric (tet mesh-based) distances are small between the center of the top and bottom of the pancake shape, while they are far using the intrinsic Laplacian distances.

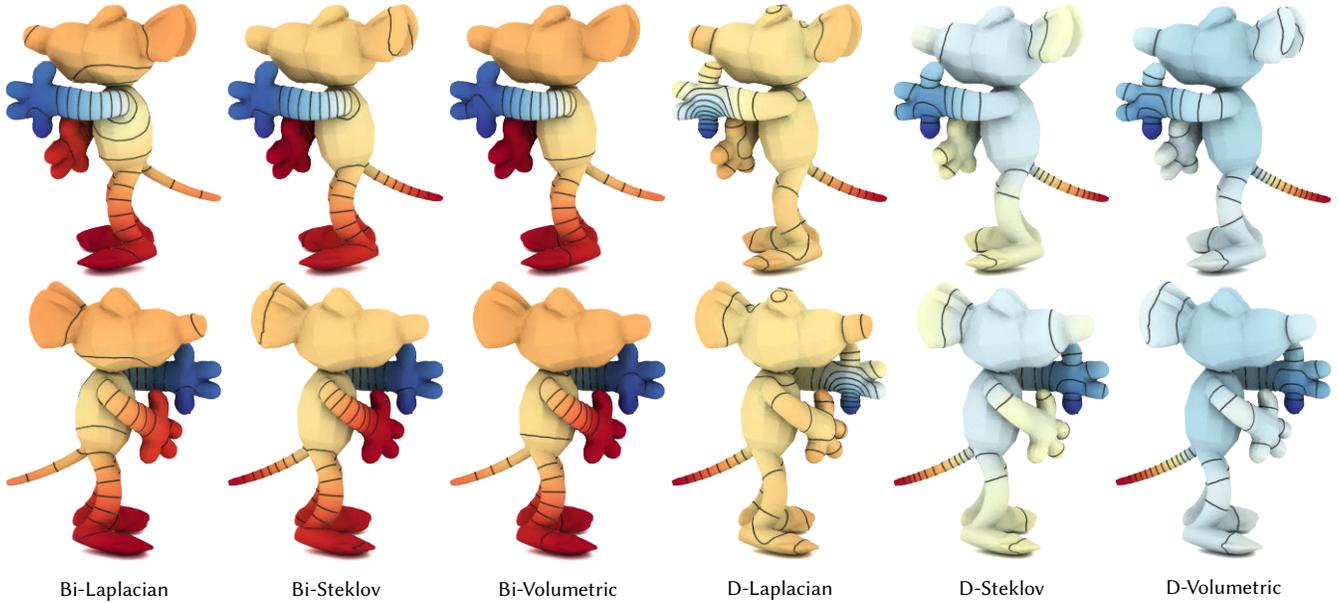

Fig. 19. Spectral and diffusion distances computed using the Laplacian, Steklov, and volumetric Laplacian spectra, illustrated using the same color scheme as Figure 18. In contrast to the example in Figure 18, while the two hands of the mouse are close in ambient space $\mathbb{R}^3$, the Steklov/volumetric distances between the hands are large because they are constrained to use paths through the *interior* of the volume.

close if you cut through the interior of the disk and far along the surface.

• MOUSE (FIGURE 19): This example shows the opposite effect. The two hands of the mouse are close in Euclidean distance, but in Steklov distance they are far. This is because our distances are *interior-aware*; unlike the disk, the shortest path between the two hands through the interior is large.

Note that computation of shortest-path distances in restricted to the interior of a non-convex triangulated surface is NP-hard [Canny and Reif 1987]. Though the Steklov family of distances is computed without interior discretization, it behaves similarly to distances computed using the volumetric Laplacian.

### 7.4 Volume-aware segmentation

Equation 2 suggests that Steklov eigenfunctions encode pointwise mean curvature, providing geometric clues for surface segmentation useful in descriptor-based algorithms [Chen et al. 2009];

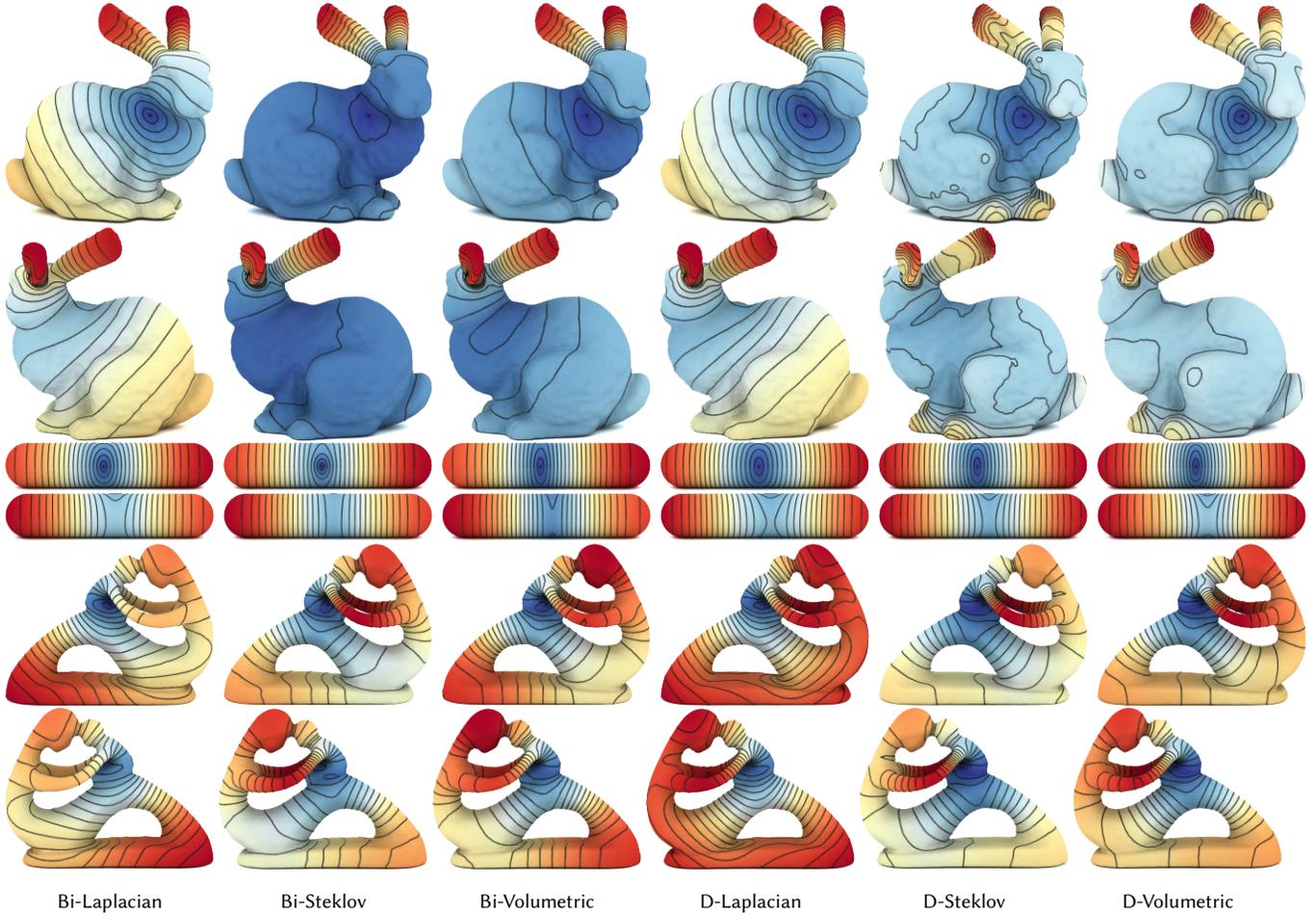

Fig. 20. Spectral and diffusion distances computed using the Laplacian, Steklov, and volumetric Laplacian spectra. For the bunny, note that the lack of surface monotonicity is the correct and expected behavior for volume distances, including the Steklov and volumetric Laplacian distances.

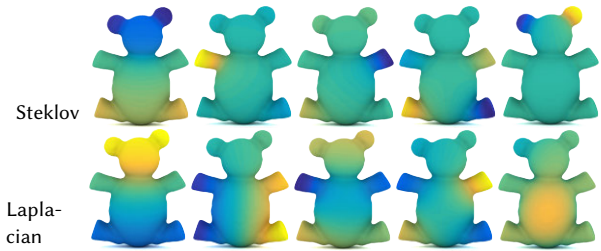

Fig. 21. The Steklov eigenfunctions corresponding to the smallest five non-zero Steklov eigenvalues, with comparison to surface Laplacian eigenfunctions. Note that the Steklov eigenfunctions align with the local extrema of mean curvature, suggesting its use as a segmentation cue.

see Figure 21 for an example. To demonstrate, in Figure 22 we apply a naïve strategy for segmentation. For the input mesh, we first compute the Steklov spectrum embedding as

$$\left( \frac{\phi_1(\mathbf{x})}{\sqrt{\lambda_1}}, \frac{\phi_2(\mathbf{x})}{\sqrt{\lambda_2}}, \ldots, \frac{\phi_k(\mathbf{x})}{\sqrt{\lambda_k}} \right),$$

where $\mathbf{x} \in \Omega$. Then, we apply the $k$-means clustering to this embedding with a user-specified number of clusters; to avoid local optima, we restart $k$-means ten times with random initialization and keep the clustering with lowest objective value. This simple strategy yields consistent segmentations: While the Laplacian embedding tends to segment the surface into flat patches, the Steklov embedding tends to segment the shape into volumetric parts.

### 7.5 Shape differences and variability

Rustamov et al. [2013] introduce a notion of shape differences based on an area-based inner product $b^a(u, v) = \langle u, v \rangle_\Gamma$ and a conformal inner product $b^c(u, v) = \langle \nabla_\Gamma u, \nabla_\Gamma v \rangle_\Gamma$ between surface functions $u$ and $v$.

Similarly, we introduce a volume-aware inner product $b^s(u, v) := \langle \nabla \mathcal{E}u, \nabla \mathcal{E}v \rangle_\Omega$. Superficially this again could be viewed as the a substitution of the Laplacian $\mathbf{L}$ used for intrinsic shape differences with the DtN operator $\mathbf{S}$. After discretization, this new inner product induces a shape difference operator $\mathbf{D} = \mathbf{S}_{\mathcal{M}}^{-1} \mathbf{F}^{\mathsf{T}} \mathbf{S}_{\mathcal{N}} \mathbf{F}$ given a functional map taking functions on the source $\mathcal{M}$ to functions on the target $\mathcal{N}$.

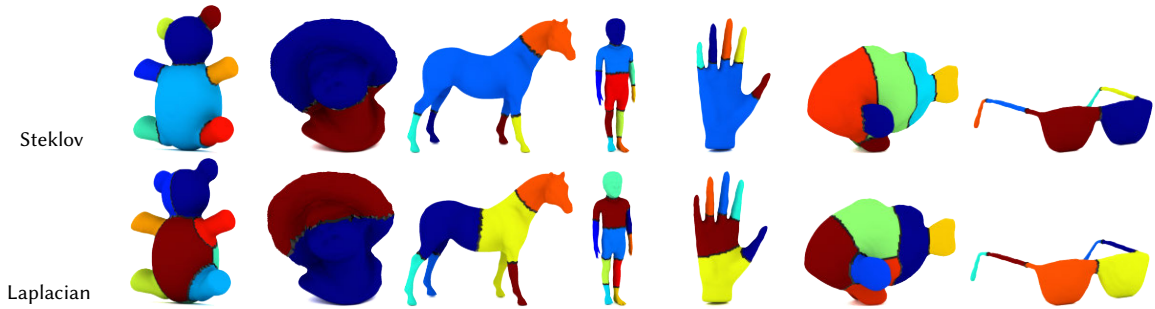

Fig. 22. Results of segmentation by the Steklov and Laplacian embeddings.

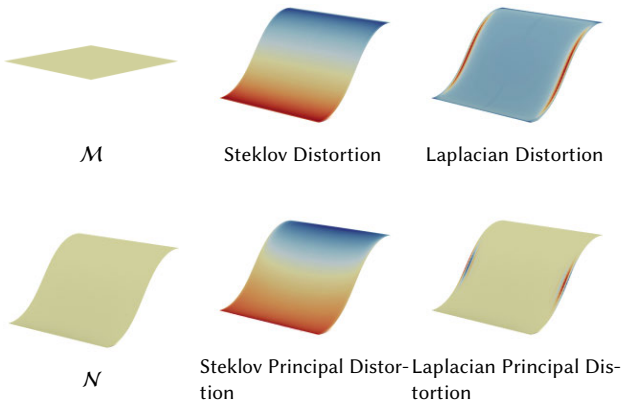

Fig. 23. Point-wise distortion, as well as principal distortion (top singular vector of shape difference operator) measured by the Steklov-based and Laplacian-based shape difference operators for a bending sheet of paper.

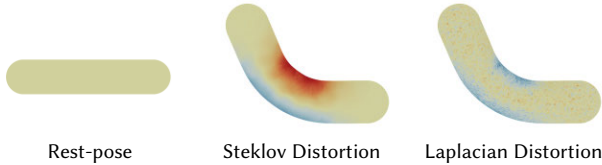

Fig. 24. Point-wise distortion measured by the Steklov-based and Laplacian-based shape difference operators. Our Steklov-based operator distinguishes bending direction of the cigar.

This shape difference operator modifies the function $f$ such that, under the map $F$ the Dirichlet energy $h(f, f)$ of $f$'s harmonic extension, will be best preserved in the sense $|b^s_{\mathcal{M}}(f, f) - b^s_{\mathcal{N}}(Ff, Ff)|$ is minimized. The action of $\mathbf{D}$ reflects how different the two geometric domains are. Inherited from the DtN operator, $\mathbf{D}$ defines a more rigid notion of shape differences that captures extrinsic shape changes not captured by intrinsic differences.

For example, Figure 23 shows a shape difference between a rectangular sheet and a bent one, Figure 24 shows a shape difference between a cigar and a bent cigar. The point-wise distortion is measured by the metric

$$(\text{distortion})_i = \frac{\mathbf{e}_i^\top \mathbf{M} \mathbf{e}_i}{\mathbf{e}_i^\top \mathbf{M} \mathbf{D} \mathbf{e}_i},$$

as in [Rustamov et al. 2013]. The Steklov-based shape difference is far less noisy and distinguishes the bending direction.

Figures 25 illustrate the effect of principal component analysis (PCA) on a collection of shape differences measured through correspondence to a base mesh; see [Rustamov et al. 2013] for details of the technique. Our new difference operator captures the extrinsic changes between surfaces that are nearly identical intrinsically, most notably the "bumpy cubes" as an extreme example. In general, our Steklov-based shape difference matrices capture more variability in the shape collections than the Laplace–Beltrami counterparts, even when using the generalized DtN operator introduced in §5.5 to study differences between open surfaces (the bending sheet and falling cloth sequences).

### 7.6 Shape exploration and retrieval

In shape retrieval applications, we may wish to distinguish the same piece of geometry in different poses, e.g. for purposes of organizing a database of human scans or for processing an animated sequence of an articulated character. In these instances, the additional discriminative features provided by DtN-based computation may be desirable. As an example, Figure 26 shows an eigenspace embedding of a collection of shapes from the TOSCA dataset [Bronstein et al. 2008] using the Steklov spectrum, similar to [Reuter et al. 2006].

### 7.7 Comparison with the Dirac operator

We next compare our method with the Dirac operator [Liu et al. 2017]. From a high level, the DtN operator is assembled using pairwise distances and relative normals, while the relative Dirac operator compares normals of nearby vertices. We highlight differences between the DtN and the Dirac operators here.

Broadly speaking, the Steklov spectrum, like the Laplacian spectrum, considers local geometric details as middle- and high-frequency geometric information; the low eigenvalues are robust (nearly invariant) to such patterns. In contrast, the low end of relative Dirac spectrum detects and discriminates such geometric details.

We provide an experiment in Figure 27 to illustrate this difference. In this example, we take a smooth bowl model and reflect the inward bump outwards to yield an isometric shape; then, we paint noise, smooth clay, and bumps with sharp edges onto the two smooth "bowls", to create four different types of geometric details. We compute both the Steklov spectrum and the Dirac spectrum on these eight shapes as shown in Figure 27a. In our experiments, the

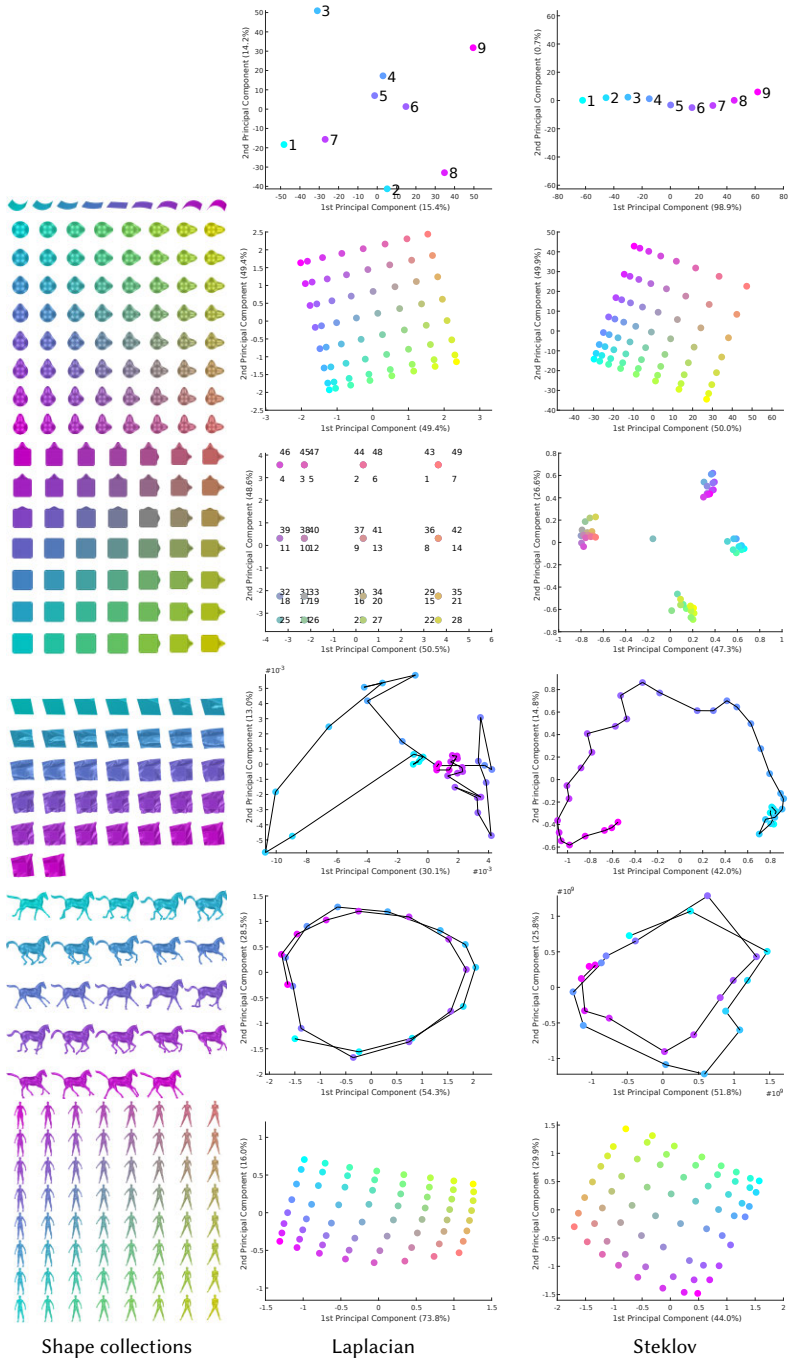

Shape collections     Laplacian     Steklov

Fig. 25. PCA on collections of shape differences. The Steklov-based shape difference reveals extrinsic deformation that conformal (Laplacian-based) shape difference fails to capture. For the bumpy cube array, the Steklov-based shape difference operator successfully classifies all shapes into four clusters: inward and inward bumps, inward and outward bumps, outward and inward bumps, as well as outward and outward bumps.

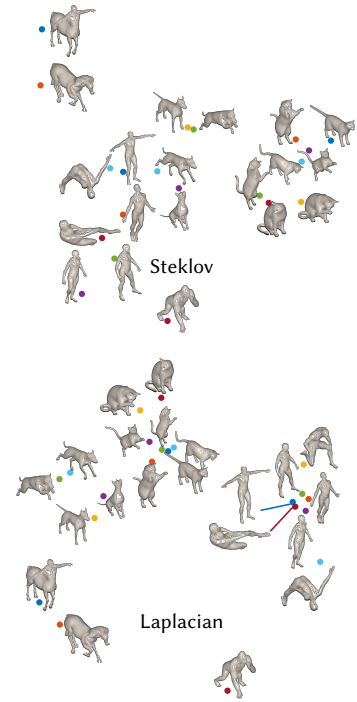

(a) 2D PCA plot of "ShapeDNAs" computed using Steklov and Laplacian eigenvalues. Shapes from the same category tend to cluster together.

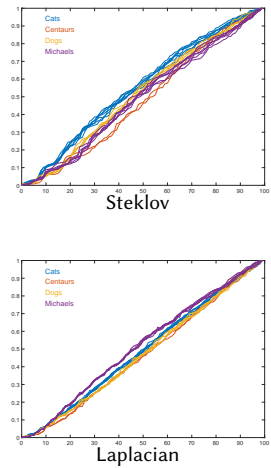

(b) For both Steklov and Laplacian eigenvalues, shapes from the same category have similar eigenvalues; the Steklov spectra cluster correctly but also distinguish between shapes within the same class.

Fig. 26: The Shape "DNA" comparision.

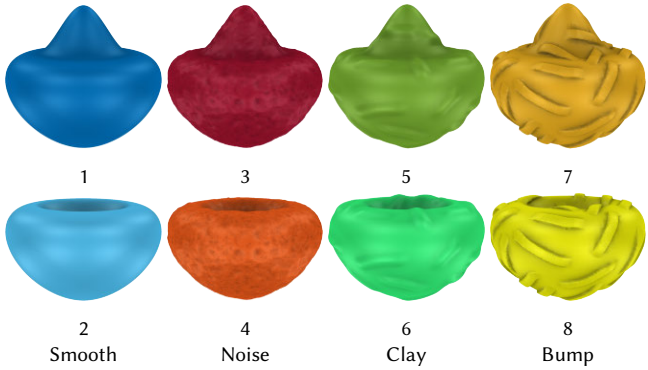

(a) Shapes to compare to [Liu et al. 2017]; the color is used to identify the shape in the following eigenvalue plots.

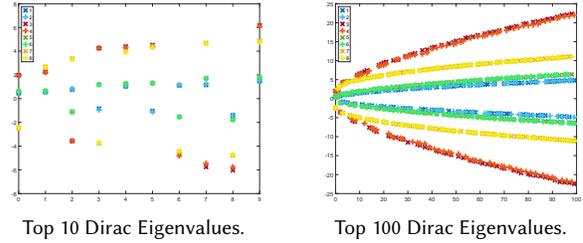

(b) The top 10 and 100 Dirac eigenvalues (which can be negative). Bowls with the same geometric details have very similar eigenvalues.

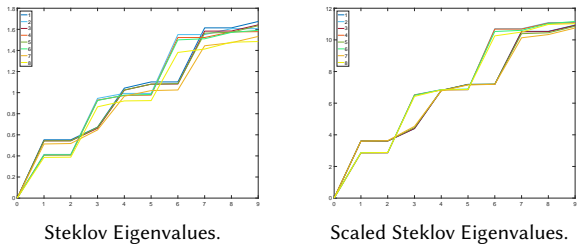

(c) The top 10 Steklov eigenvalues cluster bowls into two categories, according to whether the bump is inward or outward. Scaling the eigenvalues with the scaling factor $\mathcal{R}(\Omega)$ (defined in Appendix A) makes the separation more clear.

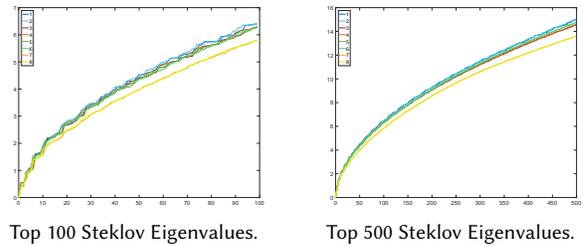

(d) The top 100 and 500 Steklov eigenvalues. The eigenvalues eventually go into four classes according to the type of geometric detail.

Fig. 27. Comparison with [Liu et al. 2017].

Dirac spectrum is computed using the source code released by the authors, which computes the spectrum of the blended operator of the relative Dirac operator (whose weight is $\sigma = 1 - 10^{-6}$) and the Laplacian operator (whose weight is $1 - \sigma = 10^{-6}$).

The top Steklov eigenvalues in our experiment roughly cluster into two categories, according to whether the bump—a large-scale volumetric feature—points inward or outward. The seperation is even more clear after we scale the Steklov eigenvalues using the isoperimetric ratio of each shape, taking into account that the scale of the shape has changed when adding geometric details. The Steklov eigenvalues gradually discriminate geometric details: The spectra of the bumpy, smooth, and noisy bowls start deviating at the $20^{\text{th}}$, $40^{\text{th}}$, and $200^{\text{th}}$ eigenvalues, resp. The principal part of the Dirac spectrum, on the other hand, discriminates the shapes into four categories, according to the type of the geometric details. Note the eight shapes are nearly isometric to each other, so adjusting the weighting $\sigma$ for operator blending does not help discriminate the direction of the bump.

The choice of using Dirac or Steklov depends on the desired behavior of the operator. The Steklov operator captures volumetric information while the Dirac operator is surface-based and accompanied by fewer guarantees. On the other hand, computing the Dirac spectrum involves sparse linear algebra, which can be cheaper than working with the dense BEM system. Geometric details, encoded at the beginning of the Dirac spectrum, are cheaper to obtain from the Dirac spectrum than from the Steklov spectrum.

## 8 CONCLUSION

The Dirichlet-to-Neumann operator provides an intuitive way to transfer the successes of intrinsic geometry processing to applications needing volumetric information. As we demonstrate using assorted applications in §7, DtN operators can substitute for Laplace–Beltrami operators *mutatis mutandis* to incorporate extrinsic geometry into existing algorithms and pipelines. The end result is an easily-applied and relatively intuitive extrinsic operator backed by theoretical understanding of its behavior on smooth domains. Our experiments further demonstrate that the boundary integral formulation of the DtN operator appears to encode meaningful geometric data on surfaces with boundary that may not enclose a volumetric region.

Our technique as it currently is implemented exhibits some limitations whose resolution will expand the applicability of our work. Most prominently, basic BEM algorithms can fail when triangle meshes self-overlap, e.g. if two triangles intersect. This drawback potentially can be resolved using a more complex implementation that detects and resolves self-overlap before evaluating boundary integrals, e.g. using techniques recently introduced to the graphics community by Zhou et al. [2016]. Another smaller drawback of our work is the assumption that volumes enclosed by surfaces are homogeneous. In physically-motivated contexts, it may be useful to incorporate anisotropy via modified Laplacian operators or by warping of the interior of the domain. This can be challenging using the boundary element method, which requires closed-form Green's functions for the differential operators involved.

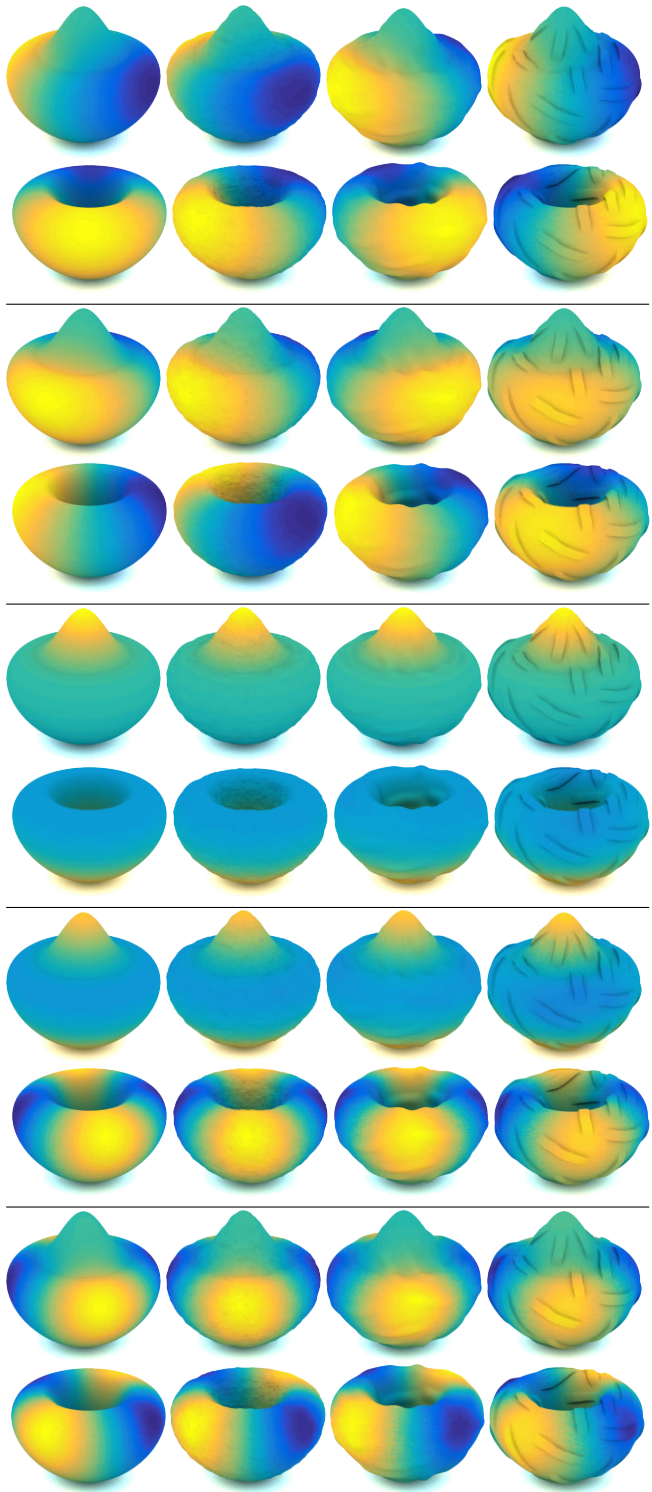

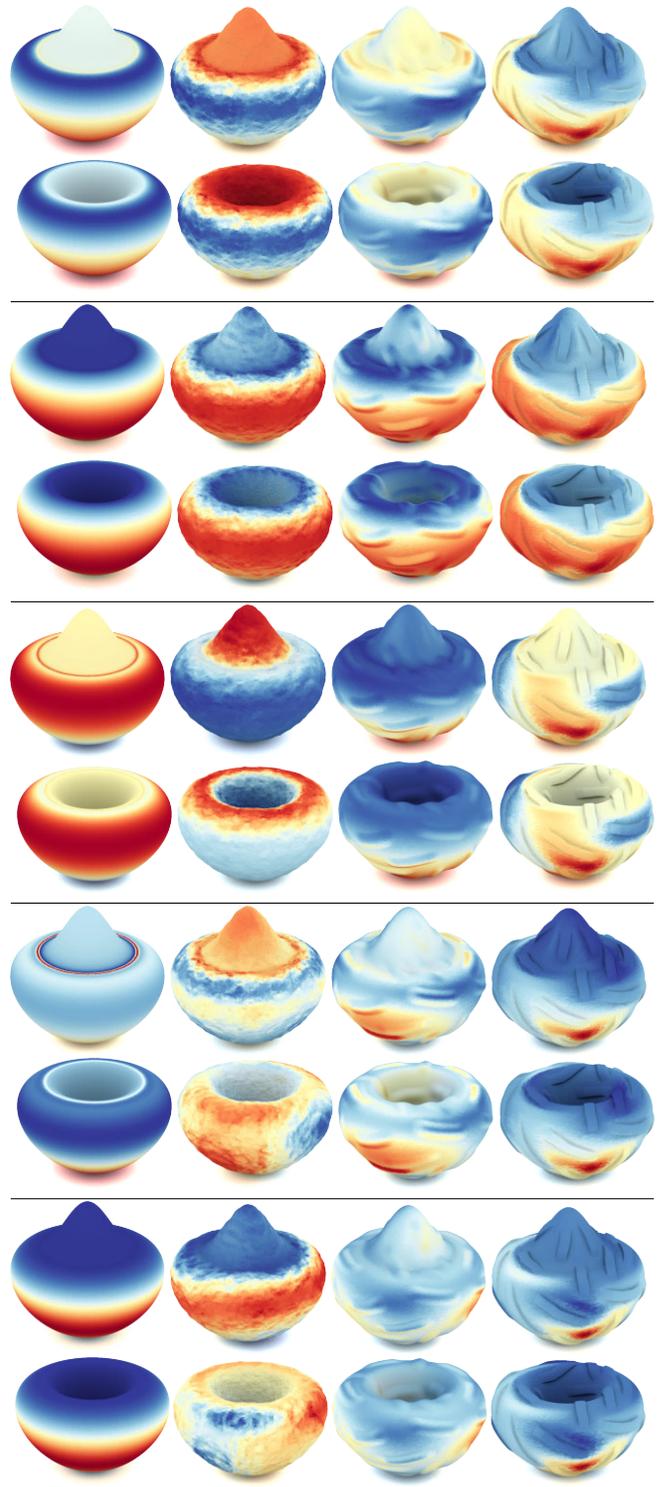

Fig. 28. Steklov eigenfunctions $2-6$ computed on the eight shapes. We skipped the first eigenfunction, which is trivially a constant function for all shapes. The eigenfunctions are nearly invariant to geometric detail, though the "rotation" within each eigenspace can be affected.

Fig. 29. Magnitudes of the first five (quaternion-valued) Dirac eigenfunctions. We can observe that the eigenfunction resonates most at nearly-flat plateaus and aligns to the clay ridges.

These potential extensions aside, future work on the computational side might focus on the efficiency of our technique. While asymptotically BEM matches if not surpasses the efficiency of FEM, our current BEM implementation could benefit from accelerated schemes for numerical quadrature and hierarchical matrix evaluation to bring this efficiency to practice. Additional speed could be gained by considering GPU-based implementations of BEM and eigenvalue algorithms, including the GPU-accelerated version of LOBPCG proposed in [Anzt et al. 2015]; their blocked matrix-vector product has some rough similarity to the hierarchical matrix method used in our BEM implementation. GPU-accelerated BEM has been considered in several recent works including [Hamada 2011; Stock and Gharakhani 2010; Takahashi and Hamada 2009; Yokota et al. 2011], a developing topic in numerical analysis that will improve the practical aspects of our method as a side effect.

More broadly, our experiments with DtN operators and the Steklov eigenvalue problem reveal the value of considering extrinsic shape in geometry processing pipelines. This and other future approaches will bring a *complete* geometric characterization to the constellation of spectral and operator-based shape processing techniques.

## ACKNOWLEDGEMENTS

The authors thank Mikhail Karpukhin for communicating the proof of Proposition 3.1, as well as Chris Wojtan and David Hahn for early conversations about BEM. J. Solomon acknowledges funding from an MIT Skoltech Seed Fund grant ("Boundary Element Methods for Shape Analysis") and from the MIT Research Support Committee ("Structured Optimization for Geometric Problems"), as well as Army Research Office grant W911NF-12-R-0011 ("Smooth Modeling of Flows on Graphs"). I. Polterovich acknowledges the support of NSERC, FRQNT, the Canada Research Chairs program, as well as the Weston Visiting Professorship program of the Weizmann Institute of Science where part of this research was performed. Meshes are courtesy of the AIM@Shape Repository, the TOSCA project, the Stanford Computer Graphics Laboratory, MIT, and the Princeton Segmentation Benchmark.

## A EIGENVALUE NORMALIZATION

For shape analysis applications where scale invariance is desired, we suggest using the following isoperimetric-ratio-based scaling factor:

$$\mathcal{R}(\Omega) := \frac{\text{Area}(\partial\Omega)}{\sqrt[3]{\text{Vol}(\Omega)}},$$

where $\text{Area}(\partial\Omega)$ and $\text{Vol}(\Omega)$ are surface area and interior volume of the domain $\Omega$. This scaling factor is justified by the following theorem

**Theorem A.1 ([Colbois et al. 2011, Theorem 1.3]).** *If $\Omega \in \mathbb{R}^3$ is conformally equivalent to a complete manifold with non-negative Ricci curvature, then $\lambda_i(\Omega) \leq c \frac{i^{2/3}}{\mathcal{R}(\Omega)}, \forall i$, where c is a constant.*

Note this is not an asymptotic formula. Figure 27 shows that this scaling factor behaves reasonably for that example, cancelling the effects of volume and area scaling.

Volume can be robustly evaluated, even for non-watertight meshes, from a boundary representation as the integral

$$\text{Vol}(\Omega) := \frac{1}{3} \int_\Gamma \mathbf{x} \cdot n(\mathbf{x}) \, d\Gamma(\mathbf{x})$$

$$= \frac{1}{18} \sum_{T_i} (\mathbf{x}_{i,1} + \mathbf{x}_{i,2} + \mathbf{x}_{i,3}) \cdot ((\mathbf{x}_{i,3} - \mathbf{x}_{i,2}) \times (\mathbf{x}_{i,2} - \mathbf{x}_{i,1})).$$

where $\mathbf{x}_{i,\{1,2,3\}}$ are positions of the three vertices in the triangle $T_i$.